\documentstyle{mn}
\begin{document}
\input BoxedEPS.tex
\SetEPSFDirectory{./}
\SetRokickiEPSFSpecial
\HideDisplacementBoxes
\title[A spectroscopic study of IRAS F10214+4724]
{A spectroscopic study of IRAS F10214+4724}
%\title[Near-infrared and optical spectroscopy of IRAS F10214+4724]
%{Near-infrared and optical spectroscopy of IRAS F10214+4724}

\author[Serjeant et al.]
{Stephen Serjeant$^{1,2}$, Steve Rawlings$^{2}$, Mark Lacy$^{2}$,
Richard G. McMahon$^3$,\vspace*{0.2cm}\\  
{\LARGE Andy Lawrence$^4$, Michael Rowan-Robinson$^1$, Matt Mountain$^5$}
\vspace*{0.3cm}\\  
$^1$ Astrophysics Group, Imperial College London, Blackett Laboratory,
Prince Consort Road, London SW7 2BZ\\
$^2$ Astrophysics, Department of Physics, Keble Road, Oxford, OX1 3RH \\
$^3$ Institute of Astronomy, The Observatories, Madingley Road,
Cambridge, CB3 0HA\\ 
$^4$ Institute for Astronomy, University of Edinburgh, Royal
Observatory, Blackford Hill, Edinburgh EH9 3HJ\\
$^5$ Gemini 8-M Telescopes Project, 950 N. Cherry Avenue, 
Tucson, Arizona, 85726, USA\\
%%$^5$ National Optical Astronomy Observatories, 950 North Cherry
%%Avenue, P.O. Box 26732, Tucson, Arizona 85726, USA\\
}
\maketitle

\begin{abstract}
The $z = 2.286$ IRAS galaxy F10214+4724 remains one of the most luminous
galaxies in the Universe, despite its gravitational lens magnification. 
We present optical and near-infrared spectra of F10214+4724, with
clear evidence for
three distinct components: lines of width
$\sim 1000 ~ \rm km s^{-1}$ 
from a Seyfert-II nucleus; $\stackrel{<}{_\sim}200$ km s$^{-1}$ 
lines which are likely to be associated with star formation; and 
a broad (\mbox{$\sim4000\rm~km~s^{-1}$}) 
C{\sc iii}]~$1909$\AA\ emission line which is
blue-shifted by \mbox{$\sim1000\rm~km~s^{-1}$} with 
respect to the Seyfert-II lines. Our study of the 
Seyfert-II component leads to several new results, including:
(i) From the double-peaked structure 
in the Ly$\alpha$ line, and the lack of Ly$\beta$, 
we argue that the Ly$\alpha$ photons have emerged through 
a neutral column of $N_{\rm H} \sim 2.5 \times 
10^{25} \rm ~m^{-2}$, possibly located
within the AGN narrow-line region as argued in several high redshift
radiogalaxies.
%although an intrinsic damped Ly$\alpha$ system with low
%gas:dust ratio cannot be excluded.
%If a background quasar were viewed through this
%gas, it would exhibit damped Ly$\alpha$ absorption.
(ii) The resonant O{\sc vi} $1032$, $1036$\AA\ doublet
(previously identified as Ly$\beta$) is in an optically thick
(\mbox{1:1}) ratio. At face value this implies an
an extreme density
($n_{e} \sim 10^{17} ~ \rm m^{-3}$) more typical of broad line region
clouds. However, we attribute this instead to the damping wings of
Ly$\beta$ from the 
resonant absorption.
(iii) A tentative detection of HeII $1086$ suggests little extinction in
the rest-frame ultraviolet.
%(iv) The physical conditions in the emission line clouds strongly
%resembles local Seyfert IIs. 
%and our interpretation of the resonant
%scattering is consistent with that in several high redshift
%radiogalaxies. 
%Nevertheless, the narrowest H$\alpha$ component is
%consistent with models with roughly equal starburst and quasar
%bolometric contributions.
%%and (iv) The detection of
%%two lines whose identifications
%%have not previously been established:
%%the first line is probably Mg{\sc vi}1806.0 which results from an 
%%ion with an ionisation potential 
%%of 141.2 eV; the second line, at a rest-frame wavelength of 
%%$\approx 2067 ~ \rm \AA$, is plausibly
%%identified with [Na {\sc v}]2067.9/2069.8.
 
\end{abstract}

\section{Introduction}
\label{sec:introduction}

The \mbox{$z=2.286$} IRAS galaxy FSC 10214+4724
is one of the most apparently luminous objects in the Universe, and 
its discovery (Rowan-Robinson et al. 1991) led to much speculation
about its possible status as a protogalaxy. This speculation was based
on the extreme  bolometric luminosity of the object (Rowan-Robinson
et al. 1991), and more specifically on the huge gas mass and
star formation rate inferred from the sub-mm molecular line and
continuum detections (e.g. Solomon, Downes \& Radford 1992; 
Rowan-Robinson et al. 1993).

This speculation was dampened by a series of papers which 
proved that F10214+4724 is being gravitationally lensed, and,
at all wavebands, is intrinsically an order of magnitude or more 
dimmer than it first appeared. Although a lensing bias was
suspected by a number of authors ({\it e.g.} Elston {\it et al.} 1994,
Trentham 1995), the first direct empirical evidence
of strong gravitational 
lensing was provided by a deep near-infrared image
(Matthews et al. 1994) which revealed an arc-like structure centred on
a galaxy close to the line of sight to F10214+4724. Several sets of authors 
published lensing interpretations (Graham \& Liu 1995; Serjeant {\it
et al.} 1995;  
Broadhurst \& Leh\'ar 1995) which were 
confirmed by the appearance of the HST image of
Eisenhardt {\it et al.} (1996): this image contained highly
elliptical, high surface  
brightness features characteristic of strong lensing, as
well as a clear counter-image. 
These papers also attempted to 
constrain the redshift of the system responsible for the
gravitational lensing, our contribution (Serjeant {\it et al.} 1995) being 
spectroscopy of two galaxies projected $\approx 1$ and $\approx$3 arcsec
from F10214+4724. This work revealed tentative $4000$\AA\ breaks 
at \mbox{$z\simeq0.90$} in both galaxies, later confirmed using
fundamental plane arguments in HST imaging (Eisenhardt {\it et al.}
1996), and also tentatively supported by 
a weak absorption line at $z=0.893$ in the 
F 10214+4724 spectroscopy of Goodrich et al. (1996).
The HST R-band image of F10214+4724 implies
magnifications of $\sim100$ 
(Eisenhardt {\it et al.} 1996) in this waveband, but
it now seems likely that differential flux magnification causes lower
magnification factors for the more
extended structure, as argued by several authors.

Gravitational lensing appeared to offer a compelling 
explanation for the extreme luminosity of FSC 10214+4724 ({\it e.g.},
Broadhurst \& L\'{e}har 1995). As a result,
the IRAS galaxy 
is no longer so extreme in its properties: 
indeed, in many respects,
it resembles local ultraluminous infrared galaxies and Seyfert-II
galaxies. 
Nevertheless, 
both Downes {\it et al.} (1995) and Green \& Rowan-Robinson (1996)
argue for a bolometric
magnification factor of $\stackrel{<}{_\sim}10$ using arguments 
based on minimum black body sizes. FSC 10214+4724 remains one of the
most intrinsically luminous objects in the Universe. 
The importance in this object still lies in studying
whether high-$z$ hyperluminous
activity, such as that seen in FSC 10214+4724, 
differs in all but scale from local objects, and in
determining the relative contributions of the starburst and AGN components
(Elston et al. 1994; Lawrence et al. 1993, 1994; Soifer et al. 1995;
Goodrich et al. 1996; Kroker et al. 1996; Hughes, Dunlop \&
Rawlings 1997).

Prior to making the observations reported in this paper
there had been no direct 
evidence for the presence of an embedded broad-line (e.g. Seyfert-I or
quasar) nucleus in $\rm F10214+4724$, although high
($\approx$20 per cent) rest-frame ultraviolet polarization
(Lawrence et al. 1993; Jannuzi et al. 1995) suggested that one was
present. This situation changed with the deep spectropolarimery of
Goodrich et al. (1996) showing clear broad lines in polarized light.
This extended the close spectral similarities between $\rm F10214+4724$ 
and Seyfert-II galaxies, specifically NGC1068, which was
first remarked on by Elston et al. (1994). 

Prior to our observations there had also been no reported 
detection of the narrow ($\stackrel{<}{_\sim}200$ km s$^{-1}$)
emission lines expected from any star-forming activity in
$\rm F10214+4724$. A star-forming component is expected if the 
analogy with NGC1068 is to be complete.
This situation also changed during the preparation of this paper. 
Using a novel imaging near-infrared spectrometer
Kroker et al. (1996) presented evidence for spatially-extended 
narrow H$\alpha$ emission just as expected if the Seyfert-II nucleus of
F10214+4724 is accompanied by a circumnuclear starburst.

In this paper we present, analyse and interpret 
optical and near-infrared spectroscopy of IRAS FSC 10214+4724.
The details of data acquisition and analysis are given in 
Section 2. In Section 3 we compare our results with previous and
contemperaneous spectroscopic studies of F10214+4724. 
In Section 4 we interpret the
data on the Seyfert-II emission line region including 
a discussion of optical depth effects on the resonance lines, and some 
modelling of the spectrum using the photoionisation code {\sc CLOUDY}
({\it e.g.} Ferland 1993, 1996). In this section we reach
conclusions about the Seyfert-II properties of 
F10214+4724 which differ significantly from those reached by previous studies. 
In Section 5 we interpret the data on the region
responsible for the narrow ($\stackrel{<}{_\sim}200 ~ \rm km s^{-1}$)
H$\alpha$ line seen in our near-infrared spectrum: this feature is likely to
be a signature of star formation. In Section 6 we pass some
concluding remarks on the nature of $\rm F10214+4724$. 
Further data on galaxies
foreground to $\rm F10214+4724$ are given in Appendix A, and our
attempts at identifying the weak emission line at $2067$\AA\ are
discussed in Appendix B.

\section{Data acquisition and analysis}
\label{sec:method}

\subsection{Optical spectroscopy}
\label{sec:opticalspectra}

We observed F10214+4724 on the nights 1995 January 28 and
1995 January 30 with the ISIS spectrograph on the William Herschel
Telescope (WHT) taking advantage of 0.7-arcsec
seeing. We refer the reader to Serjeant et al. (1995) for full 
details of these observations. The spectrophotometry is in good agreement 
with the continuum measurements tabulated by Rowan-Robinson et al. 
(1993) as well as (accounting for slit losses)
the line parameters of
Goodrich et al. (1995). There are significant disagreements between 
our measurements and those tabulated by Elston {\it et al.} (1994) ---
their line fluxes are typically larger by a factor $\sim2$ --- but these
can plausibly be attributed to imperfect correction for the 
non-photometric observing conditions experienced by Elston et al.
Neither the continuum, nor any of the emission lines 
are resolved spatially (with $\approx0.35$ arcsec pixels); 
limits on the extended line
emission are discussed by Serjeant {\it et al.} (1995). 

\subsection{Near-infrared spectroscopy}
\label{sec:NIRspectra}

We observed F10214+4724 on two occasions with the short ($150$mm)
camera and $3''$ (single pixel) slit of
the $62 \times 58$ InSb CGS4 array 
(Mountain {\it et al.} 1990) on the UKIRT. The slit was aligned
at a position angle of $90^\circ$.
The first of these observations, on
1992 February 5, used the 75 lines mm$^{-1}$ grating to obtain a
first-order spectrum centred near 1.6$\mu m$; a series of 30s
exposures were arranged in sets of four shifted by 0, 0.5, 1 and 1.5
in wavelength --- this provided Nyquist sampling and ensured that a
given wavelength was sampled by two pixels. The standard `ABBA'
nodding
pattern with a nod of 24 arcsec (8 detector rows) was used.
The total exposure time was
$9 \times 4 \times 2 \times 30\rm s$ for each point in the 126-pixel
spectrum. 
The second observation, on March 23 1992, employed the 150 lines
mm$^{-1}$
grating to obtain a second-order spectrum centred near
2.16$\mu m$. The nod in this case was $30''$ (10 rows), and the total
exposure time 10 x 4 x 2 x 40s per point. Spectra were
wavelength calibrated
using Xenon/Argon lamps and night-sky
hydroxyl lines, flux calibrated using HD105601, and the
positive and negative channels were extracted from their
respective  3-arcsec wide rows.
The wavelength calibration is accurate to $\sim0.003\mu$m in H and
$0.001\mu$m in K, and the spectral resolving powers were 250 at H
and 1400 at K.

\begin{figure}
\centering
  \ForceWidth{3.5in}
  \hSlide{-1cm}
  \BoxedEPSF{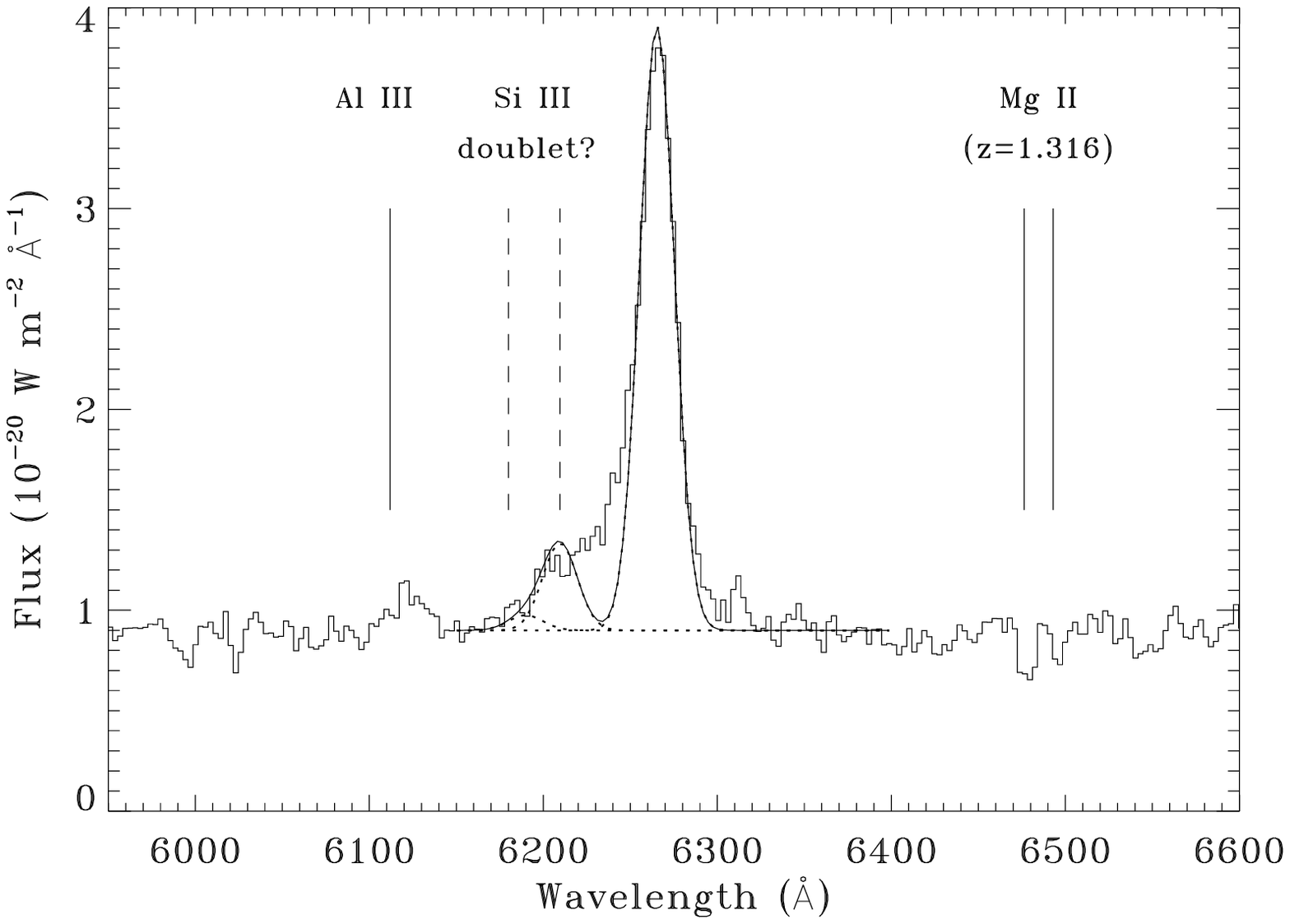}
\parbox{65mm}{
Figure 2: C{\sc iii}] line modelled as a narrow line and narrow Si{\sc
iii} doublet. The individual model components are plotted as dotted lines,
and the sum as a full line. Clearly this makes an extremely poor fit
to the data. 
The positions of the Al{\sc iii} emission line and the $z=1.361$
Mg{\sc ii} absorber are also marked.}
\end{figure}

\begin{figure}
\centering
  \ForceWidth{3.5in}
  \hSlide{-1cm}
  \BoxedEPSF{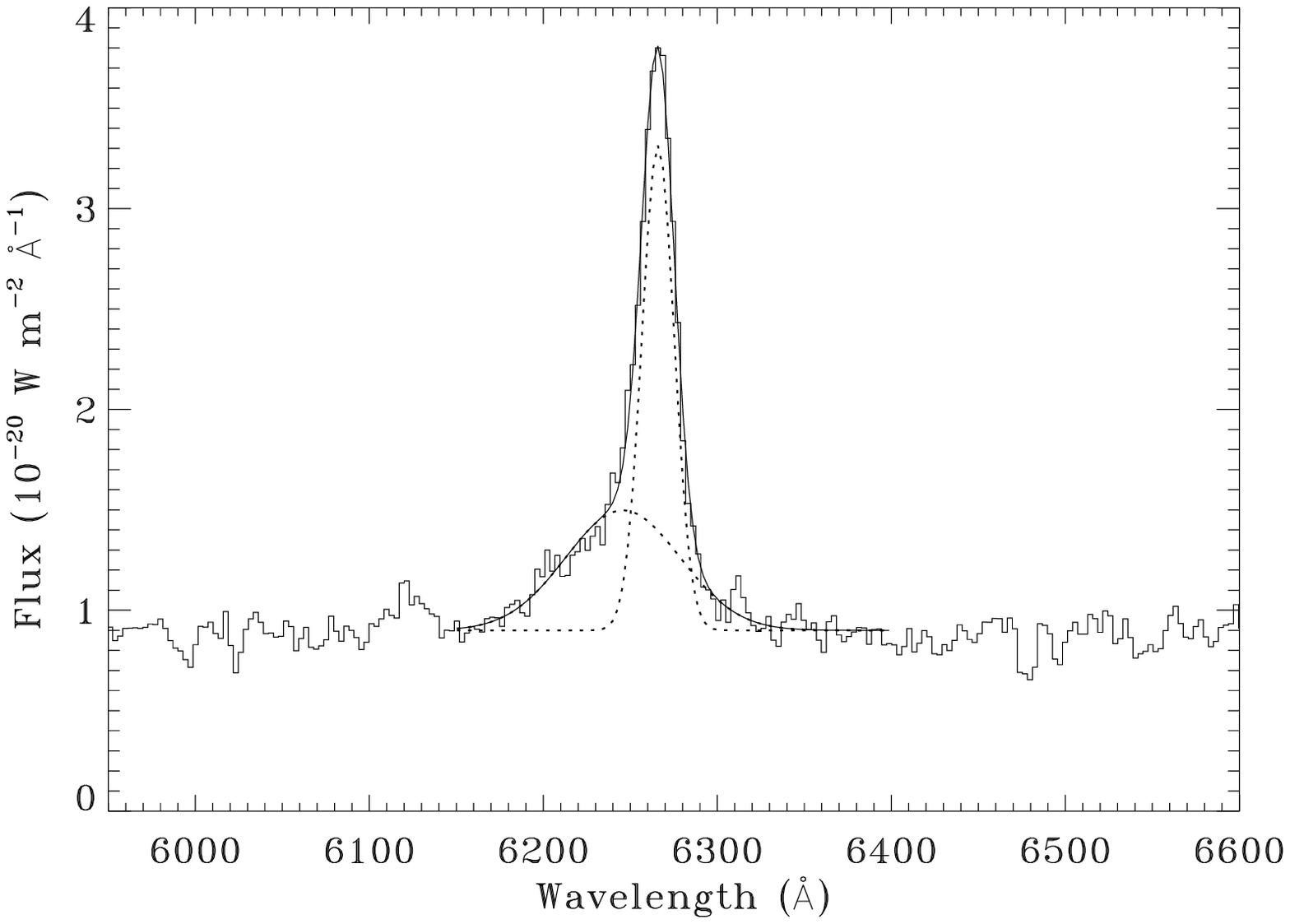}
\parbox{65mm}{
Figure 3: C{\sc iii}] line modelled as two gaussians.
}
\end{figure}

\section{Results and comparison with other spectroscopy}
\label{sec:results}

\subsection{Optical spectroscopy}
\label{sec:uvlines}

Figure~1 shows our WHT optical spectrum of F10214+4724; line fluxes, widths
and redshifts are tabulated in Table~1. Deeper spectra covering the 
region above 3900 \AA\ have been presented by Soifer et al. (1995) and
Goodrich et al. (1996), and an MMT spectrum,
comparable in sensitivity and wavelength range to our data, is
presented by Close et al. (1996); see also Rowan-Robinson et al. 
(1991) for the discovery spectrum. The velocity profiles of all
the emission lines except Ly$\alpha$ are, with FWHMs 
around $1000$ km s$^{-1}$, extremely similar. 
These line withs are small compared to quasar broad lines, but
large compared to typical AGN narrow lines which are typically
a few hundred km s$^{-1}$ ({\it e.g.} Nelson \& Whittle 1996).

The accuracy of our
wavelength calibration means that the identification of the bright 
doublet in the far blue with O{\sc vi} is secure, and we are forced 
to conclude that 
Close et al. (1996) were mistaken in preferring 
Ly$\beta$ as the identification for this feature.
This is probably the most clearly resolved example of the
O{\sc vi} doublet (see {\it e.g.} Kriss {\it et al.} 1992a,b; Laor {\it et al.}
1994), a factor we will exploit in Section 4.
Above 3900 \AA\, comparison with the published spectra shows that all 
the labelled features are real. The identification of the (rest-frame)
2470 \AA\ feature, which is also seen in NGC 1068, is new but probably 
uncontroversial. The identification of the line at
$\approx 1805$\AA\ is more uncertain: we have followed
Lacy \& Rawlings (1994) by
identifying it with Mg{\sc vi}, a line present in the {\sc cloudy} models 
discussed in section \ref{sec:cloudy}. A strong line at this wavelength
is also seen in NGC 1068 (Snijders, Netzer \& Boksenberg 1986).
We prefer the Mg{\sc vi} identification to either 
the \mbox{Si{\sc ii} $1814.0$\AA}
multiplet or the \mbox{[Ne{\sc iii}] $1814.6$\AA} line:
these were suggested as possible identifications by Snidjers et al., and
adopted for F10214+4724 by Soifer et al. 1995 and Close et al. 1996.
Note that the ionisation potential of
$Mg^{5+}$ is 141.2 eV, and thus even more extreme than the 
113.9 eV required to produce $O^{5+}$.

This leaves one unidentified line at a rest-frame wavelength of 2067 \AA.
This line is
also seen in the HST spectrum of NGC1068 (Antonucci, Hurt \& Miller
1994) where it lies in a region confused by underlying FeII multiplets 
(see also Wills, Netzer \& Wills 1980 for a discussion of a
2080 \AA\ feature in the spectra of quasars),
but in F10214+4724 it is clearly a distinct narrow
feature. Our attempts at identifying this line are discussed in
Appendix B. 

The broad asymmetric base to the C{\sc iii}]1909 line is shown in more detail
in fig~3. 
This broad base independently present (though
marginally) in both night's spectra. 
The data are reproduced by the 
sum of two Gaussian components with a broad (\mbox{$\sim 4000$ km s$^{-1}$})
component blueshifted by about \mbox{$1000$ km s$^{-1}$}
relative to a component with a similar FWHM (e.g.
$\approx 1000 ~ \rm km s^{-1}$) to the other 
optical lines. Although lines of Si{\sc III}] are expected 
to be present at some level in the blue wing of C{\sc iii}]1909
(see fig~2) we were unable to obtain as good a fit using 
a superposition of $\approx 1000 ~ \rm km s^{-1}$ lines (fig~2). 
This implies that the previous detection of 
hint of a broad C{\sc iii}]1909 line in integrated light (Goodrich
{\it et al.} 1996) is not attributable solely to the neighbouring narrow
emission lines. 
There is also an apparent blue wing to the C{\sc ii}]2326 line, but
this feature is sensitive to the correction for the atmospheric A
band. The lack of broad components to other lines may be
explainable by a combination of scattering albedo and reddening
towards the scattering surface (or between the BLR and the scatterers,
or within the scattering region itself); also lower
signal to noise may mask broad features in the  Ly$\alpha$ and N{\sc
v} region (figure 4).

The Ly$\alpha$ line has a number of unusual properties which have been
noted by other authors. First, it is weak relative to the other
ultraviolet emission lines, e.g. N{\sc v}.
This appears to reflect anomalously weak Ly$\alpha$ emission: 
ignoring Ly$\alpha$, both NGC 1068 ({\it e.g.}
Snidjers {\it et al.} 1986, Kriss {\it et al.} 1992) and the
IRAS-detected Seyfert~2 galaxy NGC 3393 (Diaz {\it et al.} 1988) have
similar ultraviolet spectra to F10214+4724.
Secondly the Ly$\alpha$ line is double peaked; this has been interpreted as 
requiring `self-absorption' (Soifer et al. 1995; Close et al. 1996). 
We will discuss these points further in Section 4.

The continuum depression shortward of Ly$\alpha$ is within the range,
though at the upper limit observed in quasars at similar redshifts
({\it e.g.} Warren {\it et al.} 1994),
consistent with the expected lack of a QSO proximity effect ({\it e.g.}
Giallongo {\it et al.} 1996 and refs. therein). 
The He{\sc ii}$1086$ line may be marginally detected. 
We confirm the presence of a Mg {\sc ii} 2800 absorption doublet
at \mbox{$z=1.316$} (see fig~2), but have insufficient sensitivity
to confirm the weaker \mbox{$z=0.892$} doublet tentatively identified
by Goodrich et al. (1996).

\begin{figure*}
\vspace*{12.0cm}
\centering
  \ForceWidth{4.0in}
  \BoxedEPSF{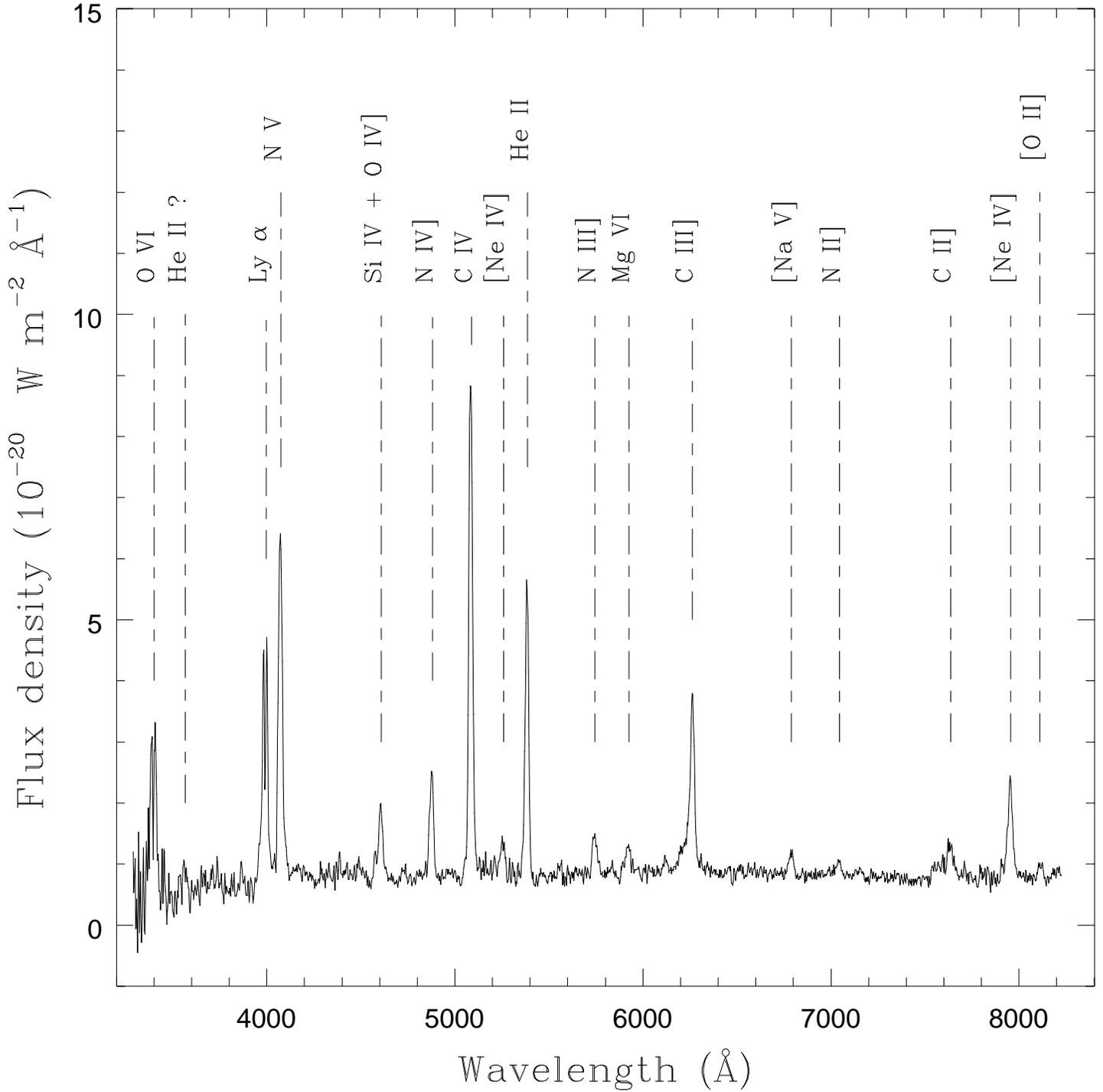}
\parbox{150mm}{
\hspace*{0cm}\vspace*{-1.5cm}\parbox{150mm}{\caption[junk]{
WHT spectrum of 
$\rm F10214+4724$. The spectrum was extracted from a full width, zero
intensity aperture, 
thereby ensuring accurate spectrophotometry. This also
included some light from a nearby companion object, particularly
redward of $\approx 7500$\AA, as discussed by
Serjeant {\it et al.} (1995). 
The emission lines are marked with their identifications.
}}
\hspace*{0cm}\vspace*{1.5cm}
}
\end{figure*}

\begin{figure*}
%\vspace*{5.0cm}
\centering
  \ForceWidth{5.0in}
%  \vSlide{-2cm}
  \TrimTop{7cm}
  \hSlide{-1.5cm}
  \BoxedEPSF{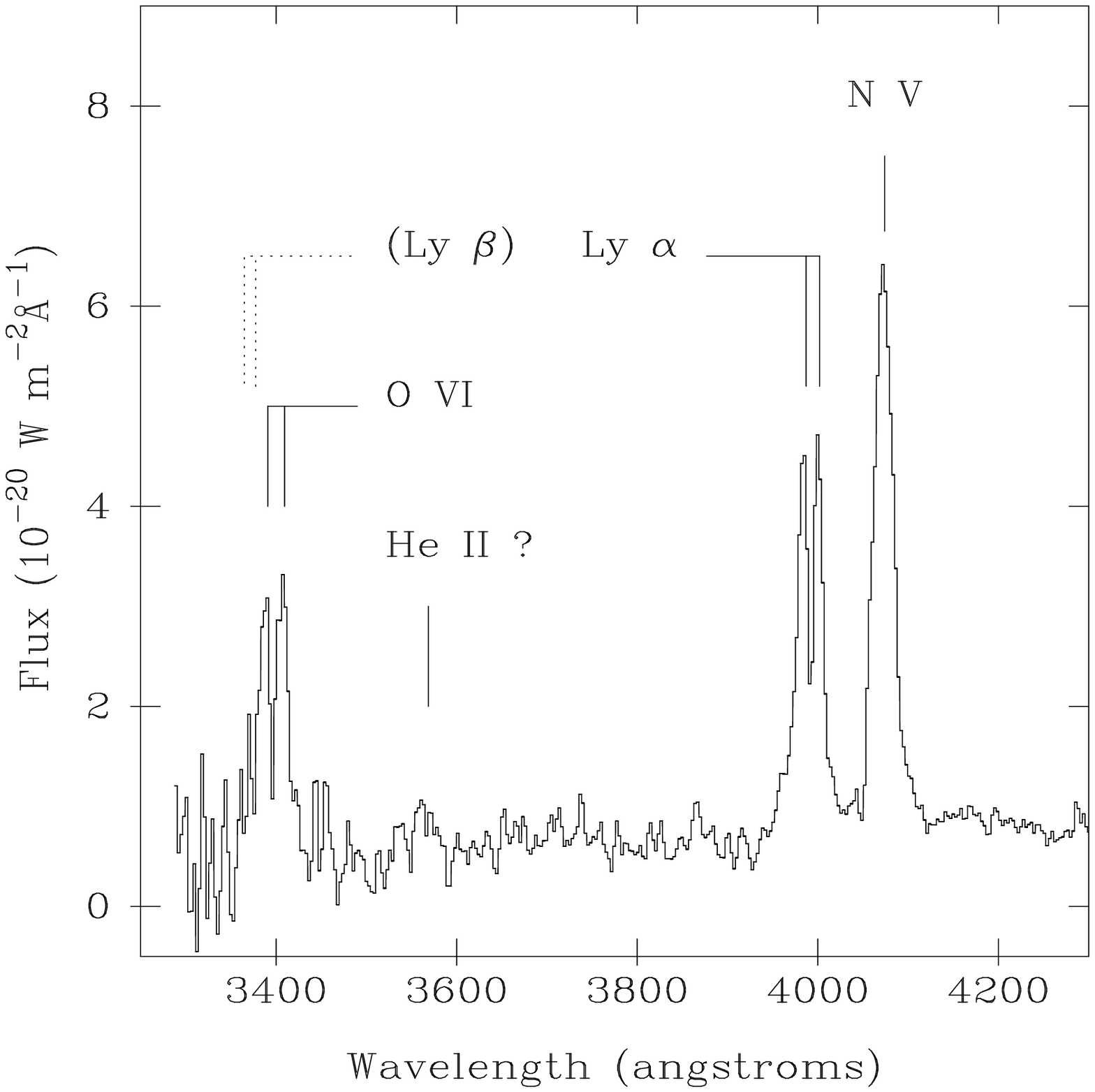}
  \parbox{150mm}{{\bf Figure 4.} Expanded UV spectrum of
F10214+4724. Positions of major emission lines are marked; also shown
are the predicted positions of Ly$\beta$ associated with the observed
Ly$\alpha$. 
}
\end{figure*}

\begin{table*}

\caption{Measurements from the WHT spectrum
of $\rm F10214+4724$.}

\begin{tabular}{llllrlll}
 & & & & & & & \\
Line & Position & Redshift  & EW & Line flux & Accuracy & Line width &
Comments \\ 

Line & /$\rm \AA$ &  & / $\rm \AA$ & 
/10$^{-19}$ W m$^{-2}$ & & $\rm km s^{-1}$ &  \\

& & & & & & \\
O VI 1031.9  & 3388 & 2.283 & 60 & 4.5 & 30\% & 1200--1600 & \\
O VI 1037.6  & 3408 & 2.284 & 60 & 4.3 & 30\% &  500--1200 & \\
CIII? 1175.7 & 3866 & 2.288 & 6  & 0.4 & 60\% &            & \\
Ly$\alpha$ 1215.7 & 3987 & 2.279 & 45 & 5.3 & 30\% & 0--900 & 
two Gaussian\\
Ly$\alpha$ 1215.7 & 4002 & 2.292 & 40 & 4.7 & 30\% & 0--600 & 
fit poor\\
N V 1240.1   & 4074 & 2.285 &130.0 & 14 & 20\%& 1500--1700 & doublet 
marginally resolved\\ 
CII 1335.3  & 4390 & 2.288 & 4 & 0.3 & 60\%  & & \\ 
Si IV 1393.7 & 4577 & 2.284 & 8 & 0.7 & 50\%  & & \\
O IV] 1402.5 & 4607 & 2.285 & 40 & 3.3 & 30\% & 1300--1500 & 
blended with Si IV 1402.8\\  
N IV] 1486.5 & 4879 & 2.282 & 45 & 4.0 & 25\% & 1250--1450 & \\
C IV 1549.0 & 5087 & 2.284 & 150 & 19.0 & 20\% & 900--1200 & \\
\mbox{[Ne IV] } 1602.0 & 5257 & 2.282 & 10 & 1.1 & 40\% & 1000-1200 & \\
HeII 1640.5 & 5385 & 2.283 & 100 & 10.0 & 20\% & 800-1150 & \\
N III] 1750 & 5743 & 2.282 & 20 & 2.0 & 30\% & 1200-1350 & \\
Mg VI 1806 & 5924 & 2.280 & 15 & 1.5 & 40\% & 1200-1350 & \\  
Al III 1857.4? & 6104 & 2.286 & 5 & 0.5 & 60\% & & Unidentified
red wing \\
C III] 1908.7 narrow & 6267 & 2.283 & 60 & $6.5$ & 30\% & 800-1000 & \\
C III] 1908.7 broad  & 6244 & 2.271 & 35 & $5.3$ & 40\% 
& $\approx$3700 & \\
\mbox{[Na V]}  2068 & 6791 & 2.265 & 10 & 1.2 & 40\% 
& 1000-1150 & see text\\
N II] 2142.8 & 7045 & 2.287 & 8 & 0.8 & 50\% & 950-1100 & 
blend with Si VII 2148 ?\\
C II] 2326.3 & 7638 & 2.284 & 30 & 3.0 & 40\% & $\approx$1500 & 
blend with OIV 2321 ?\\
\mbox{[Ne IV] } 2422.0 & 7955 & 2.284 & 40 & 5.2 & 30\% & 900-1050 & \\
\mbox{[O II]} 2470.3 & 8115 & 2.285 & 12 & 0.8 & 30\% & 1000-1150 & \\
 & &  &  & & & \\
\mbox{[O III] } 4959 & 16279 & 2.283 & 280 & 23.4 & 30\%  & $<1300$ &
\\ 
\mbox{[O III] } 5007 & 16453 & 2.286 & 580 & 49.8 & 30\%  & $<1300$ & \\
\mbox{[N II] } 6584    & 21634 & 2.286 & 14 & 6.6 & 30\%   & $<$200 & \\
H$\alpha$ & 21564 & 2.286 & 12 & 5.5 & 30\% & $<$200 & \\
H$\alpha+$[N II] broad& 21585 & $\sim2.28$  & 83 & 38.3 & 
20\% & $\sim1800$ & not deblended\\
 & &  &  & & & \\

\end{tabular} 

\parbox{150mm}{
Notes to Table 1:
Errors on the line fluxes represent $\sim$90 \%
confidence intervals expressed as a percentage of the
best line flux estimate; for the strongest lines these are dominated
by roughly equal contributions from uncertainties in fixing 
the local continuum level, and from the absolute flux calibration. 
Line widths were estimated from the FWHM of the best
Gaussian fit to each line; for the rest-frame UV lines, the range 
lower value assumes that the line-emitting 
region fills the 2 arcsec WHT slit, the higher value that it
is broadened only by the seeing.

}

\end{table*}

\subsection{Near-infrared spectroscopy}
\label{sec:Hspectrum}

Our H-band spectrum of F10214+4724 is shown in fig~5, and our K-band
spectrum in fig~6. Other near-infrared spectroscopy of F10214+4724 
has been published by Elston {\it et al.} (1994), Soifer {\it et al.}
(1995), 
Iwamuro {\it et al.} (1995) and Kroker {\it et al.} (1996).

We obtain an approximate lower limit on the [O{\sc iii}]4959+5007 to 
H$\beta$ ratio of $10$. 
Iwamuro {\it et al.} (1995) claim a tentative detection of $H\beta$ 
with an O{\sc iii} to H$\beta$ ratio of $27$, though 
the H$\beta$ is a factor of $3$ brighter than the
upper limit 
quoted by Elston {\it et al.} (1995). Iwamuro {\it et al.} used the OH
airglow supressor spectrograph on the UH 2.2m, so their detection
should be less prone to difficulties in sky subtraction. 
However, it is not clear
how imperfect sky subtraction could lead to an underestimate of
the H$\beta$ flux without a corresponding null detection of the [O{\sc
iii}] $4959$, $5007$\AA\ doublet, since the sky lines are if
anything stronger near the [O{\sc iii}] lines, and the relative
variations of the sky lines are typically $\le10\%$ (Ramsay {\it et
al.} 1992); there are also no strong atmospheric absorption lines
in the vicinity, and the spatial variation of the sky spectrum is also
expected to be negligible. 
Nevertheless, the lower resolution H~band spectrum of
Soifer {\it et al.} (1995) has a marginally detected blue wing to the
[O{\sc iii}] lines which may be H$\beta$. 

Weak features near the predicted position of the He{\sc
ii}4686 line appears to be present, which might contribute to the
unusual width of the apparent 4686 line in Soifer {\it et al.}
(1995). However, none of these features is reliably detected, and
none lies on any of the positions of $z=2.286$ emission lines expected
to be prominent (see {\it e.g.} section \ref{sec:reddening} below)
which casts doubt 
on the reliability of the apparent detections. Soifer {\it et
al.} (1995) quote a detection of He{\sc ii} $4686$\AA\ at an O{\sc
iii}/He{\sc ii} ratio of $23$, much fainter than the limit ($\sim
5$) placed in figure 5; this detection is confirmed, albeit
tentatively, by Iwamuro {\it et al.} (1995).

\begin{figure}
\centering
  \ForceWidth{0.65in}
  \TrimTop{-16cm}
  \hSlide{-4.5cm}
  \BoxedEPSF{plot_H_postref.ps}
%%%\hspace*{3cm}  %%%% comment out unless using referee style
\parbox{65mm}{{\bf Figure 5.} 
H band spectrum of F10214+4724. The position of the [O{\sc iii}] doublet
is marked, as are the predicted positions of other emission lines at
\mbox{$z=2.286$}. 
}
\end{figure}

The H$\alpha$ line profile in our CGS4 K band spectrum (figure 6)
is clearly resolved into a broader resolved component, with width
comparable to 
the UV emission lines, and a narrower unresolved component of H$\alpha$
and [N{\sc ii}]. 
We fit the K band spectrum using the stsdas.fitting.ngauss
package with the background level as a free parameter, and
with the following four gaussians: first, three unresolved (FWHM 200 km
s$^{-1}$) gaussians to model the narrower H$\alpha$ line and satellite
[N{\sc ii}] $6548, 6584$\AA\ lines, with the 6548 component
constrained to have the same redshift and $1/3$ the flux of the 6584;
second, a further 
gaussian with unconstrained amplitude, position and width. 
Although the broader component is a blend of
the broader H$\alpha$ and satellite [N{\sc ii}] lines, we model it
as a single Gaussian for simplicity. 

The results, displayed in figure 7, were very similar to a
deblending 
with 3 Gaussians of unconstrained position and width, and further
deblending models 
with alternative simplifying assumptions were attempted. 
In particular, the region between the narrower H$\alpha$ and the
narrower [N{\sc ii}]6584 lines was very poorly fit in models without a
broad excess, as might be expected from figure 7. The flux in this
region is clear evidence for an additional component. 
From the variations in the narrower line flux between the models we
estimate the flux errors in both the narrower H$\alpha$ and [N{\sc
ii}]$6584$\AA\ lines to be $\leq30\%$ (table~1). 
The flux from the broader H$\alpha$ component (distinct from
[N{\sc ii}]) is very poorly
constrained. 

Note that although the narrower
H$\alpha$ line is unresolved (FWHM less than $\sim 200$ km s$^{-1}$ at
a resolving power of $1428$), there is marginal evidence that the
[N{\sc ii}] $6584$\AA\ line is resolved. 
We found that at least some of the apparent 
excess on the [N{\sc ii}] line, if real, may be attributable to the
broader [N{\sc ii}] component. 
If we assume that
the satellite lines contribute $60\%$ of this flux, we obtain a
broader H$\alpha$ flux of $75\%$ of the total H$\alpha$. 
Note that the model does not imply the narrowest lines are
Gaussian, but rather that the K-band spectrum must have at least one
unresolved and at least one resolved component. This is exactly 
as expected for a hybrid starburst / AGN system.

\begin{figure}
\centering
  \ForceWidth{3.5in}
  \TrimTop{5cm}
  \hSlide{-1.5cm}
  \BoxedEPSF{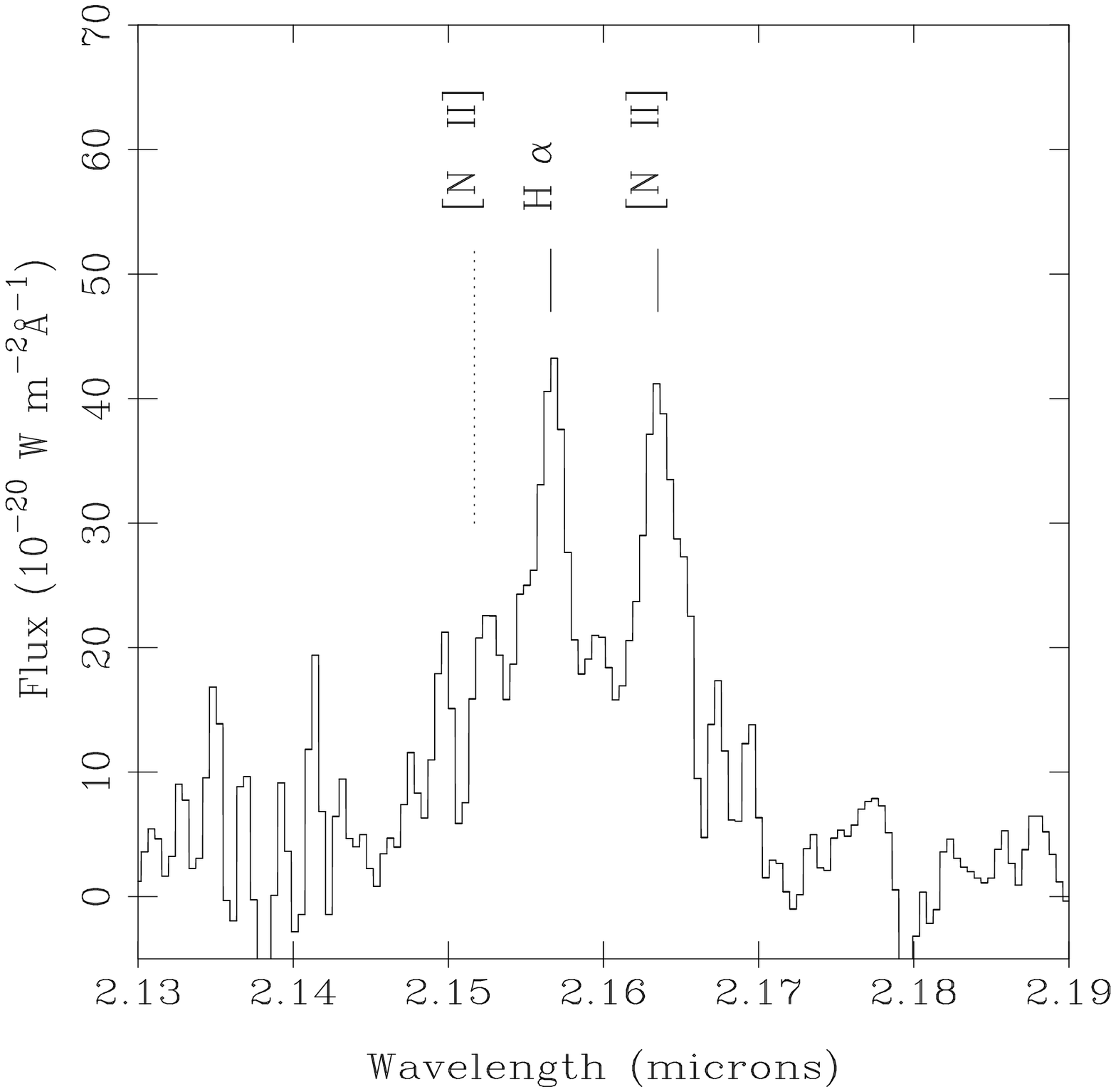}
  \parbox{65mm}{{\bf Figure 6.} 
K band spectrum of F10214+4724. The position of the narrower H$\alpha$
and [N{\sc ii}] $6584$\AA\ lines 
are marked assuming \mbox{$z=2.286$}, as is the predicted
position of the [N{\sc ii}] $6548$\AA\ line. Note the clear broad
base to these lines. 
}
\end{figure}

\begin{figure}
\centering
  \ForceWidth{3.5in}
  \TrimTop{5cm}
  \hSlide{-1.5cm}
  \BoxedEPSF{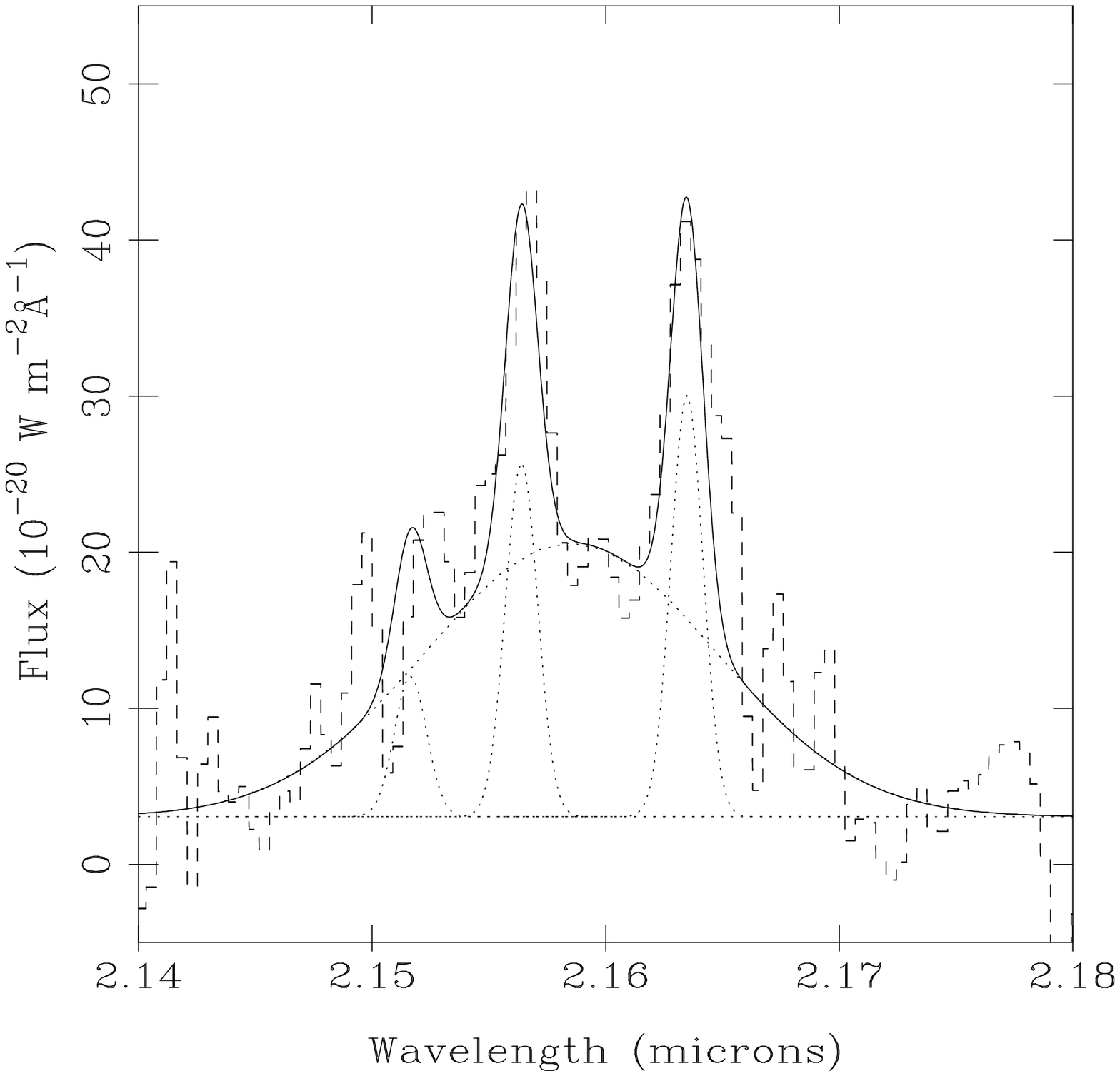}
  \parbox{65mm}{{\bf Figure 7.} 
Fit to the K band spectrum of F10214+4724 discussed in the text. 
}
\end{figure}

To summarise, we find three distinct kinematic components in our new
spectra of F10214+4724: a $\sim1000$ km s$^{-1}$ component, present 
in H$\alpha$ and in the UV lines which comes from 
the Seyfert-II Narrow-Line Region as will be discussed 
further in Section 4; a $<200$ km s$^{-1}$) component, present in
H$\alpha$, which we chose to associate with starburst activity, as will
be discussed in Section 5; and a $4000$ km s$^{-1}$ C{\sc iii}]1909 line 
which originates in a hidden Seyfert-I, or quasar, nucleus as has been 
discussed by Goodrich et al. (1996).

\section{The Seyfert-II Narrow-Line Region}

\subsection{Resonant scattering of Lyman series lines}
\label{sec:Lyalpha}

There is strong evidence for resonant scattering of the Ly$\alpha$
emission line: the profile of Ly$\alpha$ is double peaked, and the
peaks are symmetrically displaced about the predicted position of
Ly$\alpha$ at $z=2.286$ with equal flux (within the errors; table 1). 
If the Ly$\alpha$-emitting region is free of dust then the velocity 
shift $v$ of either Ly$\alpha$ peak from $1215.7$\AA is given by:
\begin{equation}
v = 195 (\frac{N_{H}}{10^{24} {\rm m}^{-2}})^{1/3} (\frac{T}{10^4
K})^{1/6}~{\rm km~s^{-1}}
\end{equation}
(Neufeld \& McKee 1988), where $N_{H}$ is the total atomic hydrogen 
column density to the source of the Ly$\alpha$ photons, 
and $T$ is the gas temperature. 
The measured velocity separation of the peaks is
\mbox{$2\times\sim560$ km s$^{-1}$}
yielding \mbox{$N_{H} = 2.5 \times10^{25} ~ \rm m^{-2}$} for
\mbox{$T = 10^4$ K}. 

The total line-centre optical depth to the source of the 
Ly$\alpha$ photons is $\tau_{\rm Ly\alpha} \sim N_{H} 
\times k_{\rm Ly\alpha}$ where $k_{\rm Ly\alpha}$ is the absorption
cross-section to the line centre of Ly$\alpha$. In the general
case of photon absorption promoting an electron upwards in energy from
level i to level j, the line-centre 
cross-section $k_{l}$ is given by
\begin{equation}
k_{l}
= \frac{1}{4\pi\epsilon_0}\frac{\sqrt{\pi}e^2 f_{ij}}{m
c \Delta \nu_{D}}
\end{equation}
where $f_{ij}$ is the oscillator strength for the transition, 
$m$ and $e$ are the electronic mass and charge, $c$ 
is the speed of light and $\epsilon_0$ is the vacuum permittivity; 
$\Delta\nu_{D} = (\frac{2 k T}{M_{A} c^2})^{0.5} \times \nu_{0}$ 
is the Doppler width of the line with
$\nu_{0}$ the central frequency,
$k$ the Boltzmann constant
and $M_{A}$ the atomic mass. In the case of the Ly$\alpha$
line, taking $T = 10^{4} ~ \rm K$, these equations yield 
$k_{\rm Ly\alpha} = 5.8 \times 10^{-18} ~ \rm m^{2}$,  
and thus $\tau_{\rm Ly\alpha} = 1.5 \times 10^{8}$.

Resonant scattering 
is extremely efficient at removing Ly$\beta$ photons even at moderate
optical depths ({\it e.g.} Netzer 1975, Osterbrock 1989), in
agreement with the null detection of Ly$\beta$ in F10214+4724.
The strong influence of resonant scattering on the observed properties of
the Lyman series lines is a major difference between the spectroscopic
properties of F10214+4724 and NGC 1068. Ultraviolet spectroscopy of the
latter (Kriss et al. 1992) shows a strong single-peaked 
Ly$\alpha$ line as well as Ly$\beta$; the observed Ly$\beta$ / Ly$\alpha$
flux ratio of 0.12 implies $\tau_{\rm Ly\alpha} \stackrel{<}{_\sim} 10$.

Is this neutral material within the narrow line region, or external,
such as a damped Ly$\alpha$ system associated with the host galaxy?
The QSO proximity effect ({\it e.g.} Giallongo {\it et al.}
1996) suggests that neutral material is
often destroyed within the ionisation cone. 
If the $N_H$ is associated with the host, it need not cause damped
absorption in F10214 itself (countering the most obvious criticism), 
because the geometry would allow scattering into the line of sight.
This neutral hydrogen could be intrinsic to the host, or could
be the result of accretion of gas-rich dwarfs; 
neutral systems with column
densities $\ge 10^{25}$ m$^{-2}$ are known to undergo a strong
cosmological evolution which dominates the evolution in $\Omega_{\rm
g}$ (Wolfe {\it et al.} 1995). 

Can the $N_H$ gas:dust ratio, derived from the Ly$\alpha$ strength,
be used to resolve this?
Ly$\alpha$ photons are conserved in resonant
scattering, 
but the Ly$\alpha$ strength can be supressed by dust absorption between
scatterings. The predicted case B Ly$\alpha$ flux 
from our broader H$\alpha$ flux implies a supression factor
of $<10^{2}$, the limit assuming no [N{\sc ii}] $6548,84$. 
In the damped Ly$\alpha$ models of Charlot \&
Fall (1991), a neutral screen with velocity dispersion $10$ km s$^{-1}$ and
column density as derived above yields a 
Ly$\alpha$ suppression of $\sim10^{4}$, but if
the gas:dust ratio is $\sim4$ times lower than the values typically
inferred from the reddening of quasars with damped systems this 
can this be reconciled with 
our observed Ly$\alpha$. 
The Ly$\alpha$ flux therefore cannot exclude the
possibility of a damped system associated with the host.
Alternatively, the same models predict
a supression $\sim10^{-1}$ for sources distributed throughout the
$N_H$ (for a velocity dispersion of $1000$ km s$^{-1}$ the supression
becomes only $\sim0.75$), and the authors argue that geometrical
factors could reduce 
this to unity. 

%This geometry dependence does however make it
%impossible to estimate the narrow line cloud gas:dust ratio.

There are difficulties with resonant scattering within the narrow line
region: the clouds in the nearer ionisation cone suffer resonant
scattering, but we nevertheless ought to have an unobscured view of the
clouds in the more distant ionisation cone. 
The contradicts {\it e.g.} the observed lack of Ly$\beta$, but this
could be explained by differential magnification of the nearer
ionisation cone, or selective obscuration 
of the more distant ionisation cone {\it e.g.} by the host galaxy.

%The lack of double peaks in other resonance lines in F10214+4724 
%({\it e.g.} C{\sc
%iv} 1549\AA) indicates that the Ly$\alpha$ scattering probably occurs
%outside the AGN narrow line region. 
%Resonant scattering from an
%extended neutral halo (even in the presence of dust) can  
%act as a Ly$\alpha$ ``mirror'' (Villar-Martin {\it et
%al.} 1996), in which case
%the IRAS galaxy could be surrounded by diffuse
%and differentially magnified Ly$\alpha$ emission. 
%This neutral hydrogen could be intrinsic to the host, or could
%be the result of accretion of gas-rich dwarfs. 
%The high column density
%is similar to damped Ly$\alpha$ systems ({\it e.g.} Wolfe {\it et al.}
%1995); 
%The nature of damped systems is controversial:
%damped Ly$\alpha$ systems are often  
%taken to be the progenitors of present-day spiral disks, 
%but some authors argue that better agreement
%with heirarchical structure formation models is obtained if
%high-redshift damped systems are comprised of irregular protogalactic
%clumps (Haehnelt {\it et al.} 1997 and refs. therein). Regardless of
%the interpretation, 
%neutral systems with column
%densities $\ge 10^{25}$ m$^{-2}$ are known to undergo a strong
%cosmological evolution which dominates the evolution in $\Omega_{\rm
%g}$ (Wolfe {\it et al.} 1995). 

We nevertheless favour models with at least some of the neutral
column in the rear of 
the narrow line clouds themselves, since the photoionisation models
below (section \ref{sec:cloudy}) favour ionisation-bounded clouds. 
Higher resolution K-band spectroscopy may resolve the AGN H$\alpha$
from the satellite [N{\sc ii}] lines, and the inferred Ly$\alpha$
supression may exclude the damped system model. 
Spatially resolved spectroscopy of F10214+4724 (while subject to the
uncertainties in the projection to the source plane) may provide a 
further confirmation:
diffuse Ly$\alpha$ could be used to trace any extended 
neutral hydrogen (Villar-Martin {\it et al.} 1996), modulo the
uncertainties in the projection to 
the source plane. The detection of non-resonance lines
spatially coincident with any extended Ly$\alpha$ would rule out
resonance scattering ``mirrors'' and identify such emission with the
AGN narrow line region or starburst knots.

\subsection{Photoionisation models}
\label{sec:cloudy}

If we exclude the resonantly scattered Ly$\alpha$, the marginally
resolved doublet N{\sc v}$1240,1243$, and the blended emission lines
(labelled in table 1), then we find no correlation of ionisation
potential with emission line width. As a result we may 
attempt to model the emission line spectrum with constant density
single slab photoionisation models, at least to first order. 

We modelled the Seyfert-II lines with the photoionisation code 
{\sc CLOUDY} (version 84.12, Ferland 1993), assuming for simplicity
that the rest 
frame UV emission lines in F10214+4724 suffer zero extinction (section
\ref{sec:reddening}). Unless otherwise stated, we used an AGN ionising
continuum as 
characterised by Matthews \& Ferland (1987) with a break at $10\mu$m
(see Ferland 1993). We considered hydrogen densities 
$8\le\log_{10}n\le 16$, where $n$ is the total hydrogen column density
({\it i.e.} molecular, neutral and ionised) in units m$^{-3}$, and
ionisation parameters $-4\le\log_{10}U\le 0$, where $U$ is the
dimensionless ratio of incident ionising photons to hydrogen density,
{\it i.e} $U=\Phi(H)(nc)$. Here $\Phi$ is the surface flux of ionising
photons in m$^{-2}$s$^{-1}$, and $c$ the speed of light. The
calculation was stopped at a column density of $10^{27}$ m$^{-2}$, or
at a temperature of $4000$K, since below this temperature the emission
line flux is negligable. We used the default solar metallicity (see
Ferland 1993), and {\sc CLOUDY} assumes a plane-parallel geometry. 

The He{\sc ii}:Ly$\alpha$:O{\sc vi}:C{\sc iv} ratios are sensitive to
the UV-soft Xray continuum ({\it e.g.} Krolik \& Kallman 1988), so are
sensitive to the presence or absence of an active nucleus. A black
body ionising continuum at $40,000$K, resembling the continuum from
the hottest OB stars, as severe difficulty in producing O{\sc
vi}$1032,1037$\AA\ or Mg{\sc vi} $1806$\AA, due to the extremely high
ionisation potentials of O$^{4+}$ and Mg$^{4+}$ ($.11$ and $.14$ keV
respectively). The AGN ionising continuum defined above had no
difficulty in reproducing these lines. We are therefore confident that
the $\sim 1000$ km s$^{-1}$ emission line region is predominantly
photoionised by the active nucleus. 

None of the single-slab models provided an adequate fit to all the
rest-frame UV emission lines. 
We found that similar, but not identical, conditions
($\log_{10}n\sim10$, $\log_{10}U\sim-1.5$) were implied by
emission lines [Ne{\sc iv}] $1602$, He{\sc ii} $1640$, N{\sc iv}] $1486$,
N{\sc iii}] $1750$, Ne{\sc v} $3426$, Ne{\sc iii} $3869$. 
For example, the largest [Ne{\sc
iv}] $2424$ / He{\sc ii} $1640$ ratio in our models was with a density
and ionisation parameter of $\log_{10}=10$, $\log_{10}U=-1.4$. This
underpredicts the O{\sc vi} and Mg{\sc vi} lines, as well as the N
lines. However, 
the N{\sc iii}] $1486$, $1750$ and N{\sc
v}$1240$ ratios are all consistent with $\log_{10}U\sim -1.25$; this
self-consistency of the N lines suggests that the strength of the N
$1240$\AA\  line may be due mainly to an abundance effect, as also
argued in the IRAS detected Seyfert II NGC 3393 (Diaz {\it et al.}
1988) rather than due to differential magnification (Broadhurst \&
L\'{e}har 1995). If so, this would further reduce the proportion of
H$\alpha$ in the broader component of figure 7. 

This model reproduced the C{\sc iv}:C{\sc iii}] ratio, but
overpredicted their strength by $\sim\times 5$. A low carbon abundance
($\sim 0.5\times$ solar) might resolve the anomaly, perhaps caused by
depletion onto dust grains. Much more problematic is the lack of an
[O{\sc iii}] $1665$\AA\ emission line, which in these conditions
should have a flux $\sim 10\%$ of He{\sc ii} $1640$\AA. There does not
appear to be any simple solution; we note however that an identical
problem was found by Kriss {\it et al.} (1992b) in NGC1068, who argued
that shock-heated gas would also produce the [O{\sc iii}] $1665$
line.

Finally, we can use our limits on the density and ionisation parameter
to derive the sizes and masses of the NLR clouds.
The clouds are
ionisation bounded 
in the conditions we derived from the emission line ratios, 
as is also conventionally
assumed for AGN narrow line regions, and unlike {\it e.g.} the
matter-bounded 
extended emission line region models of Wilson {\it et al.} 1997. This  
supports our model for the resonant scattering (section \ref{sec:Lyalpha}),
and is also consistent with the presence of [O{\sc i}] in the Keck
spectrum (Soifer {\it et al.} 1995). 
Assuming the absorbing column does indeed occur within the NLR clouds
(assumed 
constant density for simplicity), we obtain a
characteristic size scale of
\begin{equation}
\begin{array}{lclcll}
S   & = & N_H/n_H & = & 2.5\times10^{-2 \pm 0.5} {\rm pc} & \\
\end{array}
\end{equation}
where the errors represent the acceptable range in the models, rather
than {\it e.g.} gaussian noise. 
The mean mass of the NLR clouds is given by
\begin{equation}
\begin{array}{lcll}
M   & = & u\times n_H \times S^3 & = 1\times 10^{-2\pm 1} {\rm M}_\odot \\
%    & = & 1.6606\times 10^{-27} {\rm ~kg~} \times 10^{10.5 \pm 0.5} & \\
%    &   & \times (2.5\times 10^{14.5 \pm 0.5})^3 & \\
%    & = & 1.3\times 10^{-2\pm 1} {\rm M}_\odot & \\
\end{array}
\end{equation}

Unfortunately it is not possible to derive a covering factor
or volume 
filling factor, without spatially resolved spectroscopy. Briefly, $N$
clouds with covering
factor $C$ at mean distance $D$ parsecs from a luminosity $L$ quasar,
are indistinguishable from $N,C/4,2D,4L$. We can however estimate the
number of narrow line clouds, and the total cloud mass containted in the
narrow line region. Using {\sc cloudy}, we find the emissivity of a
single cloud of density 
$10^{10.5\pm0.5}$ m$^{-3}$ with ionisation parameter $10^{-1.25\pm0.25}$
in {\it e.g.} the He{\sc ii} $1640$ line is $\epsilon=10^{-2.4\pm0.7}$ W
m$^{-2}$. For a 
magnification factor of $10{\cal M}_{10}$, our observed line flux implies
the number of narrow line 
clouds is
\begin{equation}
\begin{array}{lcll}
N   & = & (10^{-18}{\rm W~m^{-2}} \times 4\pi D_L^2) / (4\pi\epsilon S^2\times 10{\cal M}_{10}) & \\
    & = & 1\times 10^{7\pm 1.4}{\cal M}_{10}^{-1}h_{50}^{-2} & \\
%%%%    & = & 5\times 10^{9\pm 1.4}{\cal M}_{10}^{-1}h_{50}^{-2} & \\
\end{array}
\end{equation}
where $D_L$ is the luminosity distance to F10214+4724, given (assuming
$\Omega_0=1$, $\Lambda=0$, $H_0=50h_{50}$ km s$^{-1}$ Mpc$^{-1}$) by
\begin{equation}
D_L =  0.04ch_{50}^{-1}(1-(1+2.286)^{-0.5})(1+2.286)
\end{equation}
(We also note that variations in $S$ correlate with
variations in the cloud emissivity, so the ``errors'' do not simply
add). The inferred total mass of ionised gas within 
the narrow line region is consistent with estimates in local AGN
({\it e.g.} Osterbrock 1993):
\begin{equation}
\begin{array}{lcll}
NM  & = & 10^{-18} 4\pi D_L^2  u n_H S^3/ (4\pi S^2\epsilon 10{\cal M}_{10}) & \\
    & = & N_H \times 5\times10^{12\pm 0.7}{\rm m}^2~{\cal M}_{10}^{-1}h_{50}^{-2} {\rm ~kg} & \\
    & = & 2\times10^{5\pm 0.7}{\cal M}_{10}^{-1}h_{50}^{-2} M_\odot & \\
%%%%%    & = & 6\times10^{7\pm 0.7}{\cal M}_{10}^{-1}h_{50}^{-2} M_\odot & \\
\end{array}
\end{equation}
 Our choice of He{\sc ii}, while free of
metallicity effects, assumes negligable extinction
(section \ref{sec:reddening}) which will be testable with higher quality
infrared spectra.
Note that if
spatially resolved 
spectroscopy 
detects the emission line counterimage, it may be possible to
determine ${\cal M}_{10}$  independently of spatial structure in the
IRAS galaxy. As a 
corrolary to the covering factor 
calculation one may then also obtain a robust estimate of the
luminosity of the 
obscured quasar (Goodrich {\it et al.} 1996).

\subsection{Optically-thick O{\sc vi} emission}
\label{sec:Ovi}

The resonant O{\sc vi} 1034, N{\sc v} 1240 and
C{\sc iv} 1549 lines are doublets in the isoelectronic lithium
sequence. If such lines emanate from a region which is 
optically-thin to resonant scattering, then their
doublet line ratios are given by the ratio of the statistical weights
of the upper levels, {\it i.e.} \mbox{2:1} with the 
lower wavelength line of the doublet the brighter.
However, in optically-thick regions, photon trapping
and collision de-excitation can cause the levels to thermalize so that the
flux ratio approaches \mbox{$\sim$1:1}; this thermalization
occurs only at high values of optical depth and 
electron density $n_{e}$, namely when 
$n_{e} \tau_{0} \stackrel{>}{_\sim} 10^{22} ~ \rm m^{-3}$
(Hamann et al. 1995). A doublet ratio of 1:1 therefore implies extreme
densities and/or optical depths. 

The O{\sc vi} \mbox{$1031.9, 1037.6$} doublet in F10214+4724 is
clearly resolved (Fig. 4), and its flux ratio \mbox{1:1} 
suggests line thermalization. We first consider 
whether this ratio is strongly influenced by
blanketing by the Ly$\alpha$ forest lines --- both a simple empirical
test and a brief analytical argument suggest it is not. The test 
involved fitting low-order polynomials to the UV continuum
longward of Ly$\alpha$, and extrapolating these fits to shorter
wavelengths to estimate the level of line blanketing. We then
superimposed a model \mbox{2:1} doublet at several positions on this
absorption spectrum and found it impossible to obtain an apparent 
\mbox{1:1} ratio. This is in good agreement with
the following analytical argument. 
The number of Ly$\alpha$ forest lines per unit redshift per unit
rest-frame equivalent width is well fitted by an expression of the form
\begin{equation}
\frac{d^2 N}{dz dW} = \frac{A_0}{W^*}(1+z)^\gamma \exp(-\frac{W}{W^*})
\end{equation}
where $\gamma\simeq1.89$, $W^*\simeq0.27$\AA\ and $A_0\simeq10$ ({\it
e.g.} Bechtold 1994). By integrating this over the redshift range
$1.770$ to $1.792$ ({\it i.e.} in the neighbourhood of the doublet and
exactly encompasssing the width of either line of the 
doublet), and further integrating $dN/dW$, 
it can be shown that a maximum of two Ly$\alpha$ forest lines are
expected to lie on the position of one O{\sc vi} line. 
In order for a \mbox{2:1} doublet
to be suppressed to \mbox{1:1}, the line blanketing over the shorter
wavelength line must be 
approximately a factor of $2$ (in addition to the mean 
blanketing); the probability of this occuring is negligibly small,
a conclusion which is insensitive to the uncertainties in 
$\gamma$, $W^*$ and $A_0$. 
\begin{figure*}
\centering
  \ForceWidth{5.0in}
  \BoxedEPSF{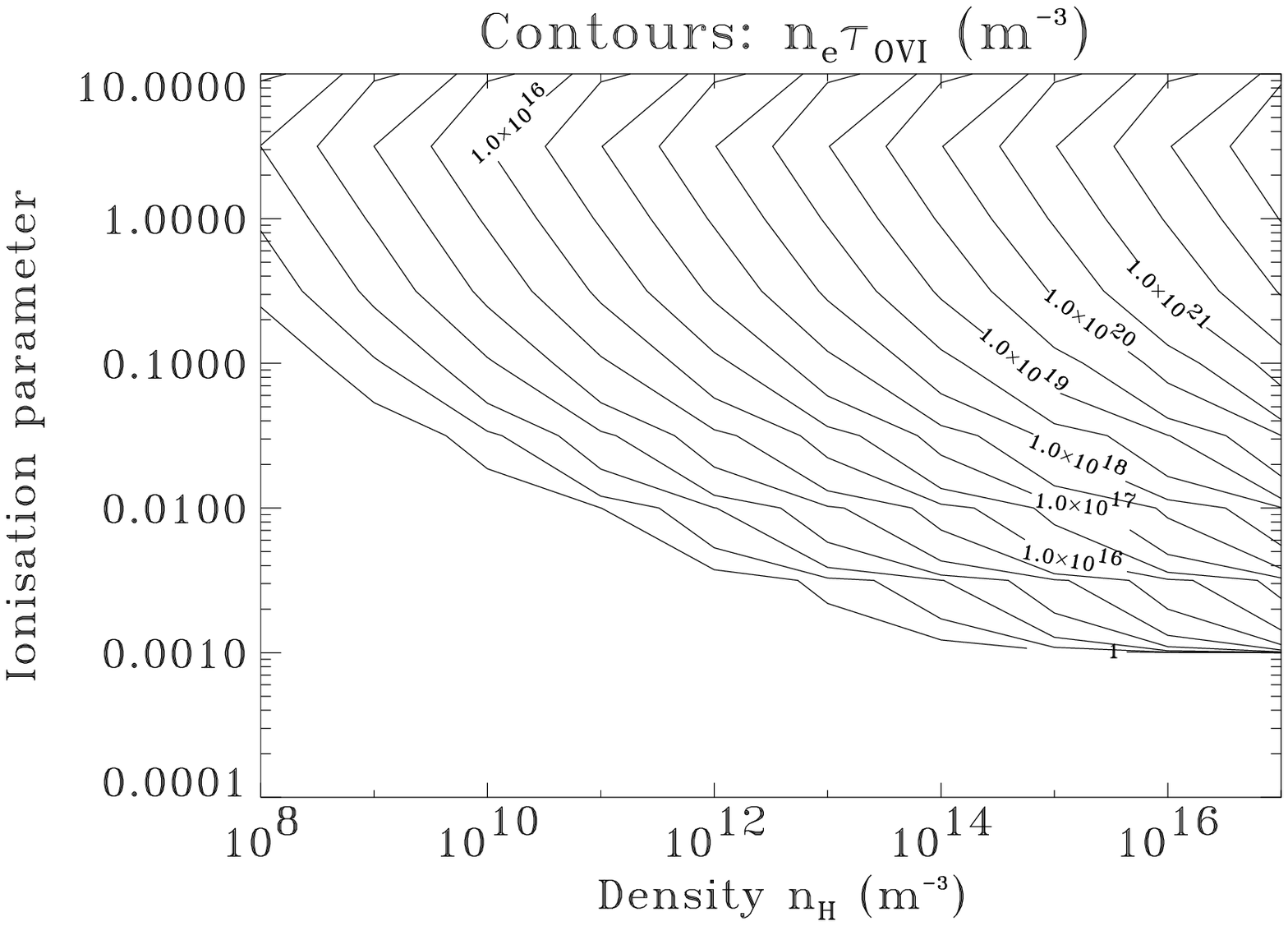}
  \parbox{150mm}{{\bf Figure 8.} 
Results of the photoionisation model discussed in the text. 
The contours show the product of the $n_{\rm p}$-weighted average
electron density and
O{\sc vi} $1032+1037$\AA\ optical depth, $n_{\rm e}\times\tau_{\rm
O{\sc vi}}$, 
as functions of ionisation parameter and density. (Almost identical results
were obtained replacing this $n_{\rm e}$ with 
the innermost zone electron density, the outermost, or the total
Hydrogen density.) The contours are 
spaced logarithmically in steps of $0.5$. Only where $n_{\rm
e}\times\tau_{\rm O{\sc vi}} \stackrel{>}{_\sim} 10^{22}$ m$^{-3}$ does the
O{\sc vi} doublet form a 1:1 ratio.}
\end{figure*}

This seems to imply that the O{\sc vi} emitting region 
in F10214+4724 is both optically thick 
($\tau_{\rm O{\sc vi}} \gg 1$) and dense enough for the 
doublet to have become thermalised, {\it i.e.} $n_{\rm e}\tau_{\rm
O{\sc vi}} \stackrel{>}{_\sim}10^{22}$ m$^{-3}$. 
We will next show in the
that the optical depth $\tau_{\rm O{\sc vi}}$ is unlikely
to exceed $10^6$, resulting in apparent broad-line-region-like
conditions for 
the narrow O{\sc vi} emitting region.

%Since each line in the 
%doublet has single rather than double-peaked structure (c.f. Ly$\alpha$
%discussed in section \ref{sec:Lyalpha}), limits on the amount of
%wavelength diffusion 
%suggest an upper limit to the optical depth of
%$\tau_{\rm O{\sc vi}} < 10^{8}$, and hence a lower limit to the
%density of the region of $n_{e} > 10^{14} ~ \rm m^{-3}$.
%Using equation X to estimate $k_{O{\sc vi}} = 6.5 \times 10^{-18}
%~ \rm m^{-2}$, and assuming a solar \mbox{O:H} ratio, the
%associated column density is $N_{H}^{*} \sim 2.5 \times 10^{20} 
%\tau_{\rm O{\sc vi}} ~ \rm m^{-2}$. Since O{\sc vi} is formed
%and resonantly scattered only within the zone
%containing $\rm O^{5+}$, and since Ly$\alpha$ can be scattered 
%even in neutral zones where no line emission is produced, it seems likely 
%that $N_{H}^{*}$ can approach but not exceed the value of $N_{H}$ 
%estimated in Section \ref{sec:Lyalpha} (assuming that the bulk of the
%Ly$\alpha$ photons are generated in the highly-ionized region). 
%This implies $\tau_{\rm O{\sc vi}} \sim 10^{5}$ and 
%$n_{e} \sim 10^{17} ~ \rm m^{-3}$.

The latest release of {\sc CLOUDY} (version 90.03a, Ferland 1996)
incorporates the O{\sc vi} $1032$,$1037$\AA\ doublet lines separately,
so we used this code to explore the doublet ratios and O{\sc vi}
optical depths for a wide range of physical conditions. We used the
same AGN ionising continuum, metallicity and stopping conditions as
above. The results are shown in fig.~8. We find that the 1:1 ratio,
also where the 
$n_{\rm e}\tau_0\stackrel{>}{_\sim}10^{22}$ m$^{-3}$
limit applies, is possible only with hydrogen densities in excess of
$n_{\rm H}\simeq 10^{17}$ m$^{-3}$. This conclusion was found to apply
equally to $N_{\rm H}=10^{26}$ m$^{-2}$ column and/or $Z=10Z_\odot$
metallicity gas. 
Such a high value of $n_{\rm H}$ is characteristic of the broad-line region 
of AGN (e.g. Ferland et al. 1992) rather than the lower values
($n_{e} \sim 10^{11} ~ \rm m^{-3}$) found in classical narrow-line regions.

Such a radical conclusion can be avoided, albeit with some fine
tuning, by considering the Ly$\beta$ resonant absorption.
This absorption is likely to be saturated, so the O{\sc vi} doublet
may lie inside its damping wings. Using the $N_H$ value derived above,
we obtain an optical depth of about $0.3$ for the $1032$\AA\ line and
$0.09$ for
the $1038$\AA\ line ({\it i.e.} a factor $\sim3.6$). To obtain a
\mbox{$1:1$}
ratio from \mbox{$2:1$}, one needs an optical depth of around $1$ to the
$1032$\AA\ line, close to the estimate from the observed $N_H$. This
explanation avoids the high densities, but is sensitive to the assumed
neutral column, and furthermore takes no account of geometrical
effects (Vilar-Martin {\it et al.} 1996). 

%At such densities collisional de-excitation will eliminate all the
%forbidden lines, and even semi-forbidden lines like CIII]1909. 
%We suggest, therefore, that the O{\sc vi} lines (and possibly some of
%the observed flux from other high-ionisation 
%lines, such as N{\sc v} 1240) form in a region akin to the broad-line
%region, 
%whereas lines like CIII]1909 
%are from regions further from the exciting source.
%We might also expect to see
%\mbox{$\sim$1:1} flux ratios for other resonance doublets: unfortunately the
%separations of the N{\sc v} and C{\sc iv} doublets are insufficient to
%be well resolved because of the large intrinsic widths of the lines, and
%because of the poor spectral resolution of our observations. 

Inspection of the ultraviolet spectrum of NGC1068 (Kriss et al. 1992)
suggests that the O{\sc vi} doublet is again in a \mbox{$\sim$1:1} flux ratio.
However, there is no evidence of resonant scattering in this object.

\subsection{Reddening of the Seyfert-II nucleus}
\label{sec:reddening}

We can place limits on the extinction in the observed UV
from the He{\sc ii} $1640$\AA/He{\sc ii} $1086$\AA\ 
ratio, which in the absence of reddening is predicted to
be~7 ({\it e.g.} Seaton 1978). Although the signal to noise shortward of
Ly$\alpha$ in Fig.~4 is poor, and despite possible blanketing by
Ly$\alpha$ forest lines, the He{\sc ii} $1086$\AA\ line is
marginally detected at \mbox{$(\sim7\pm50\%)^{-1}$} times the flux of
the $1640$\AA\ counterpart. If real, this suggests an extinction of
between zero and $0.75$ magnitudes around $1200$\AA\ in the rest frame
of F10214+4724 ({\it i.e.} \mbox{$A_{\rm V}<0.2$}), or $2600$\AA\ in
that of the lens ({\it i.e.} \mbox{$A_{\rm V}<0.4$}). 
We can also obtain a rough extinction estimate from the
H$\alpha$:He{\sc ii}$ 1640$\AA\ ratio, predicted to be about $2.3$ in
our solar metallicity models above. We find 
(H$\alpha$+[N{\sc ii}]):He{\sc ii}$=3.8$, 
consistent
with an H$\alpha$:N{\sc ii} ratio resembling {\it
e.g.} NGC 1068 ({\it e.g.} Bland-Hawthorn {\it et al.} 1991) and zero
extinction.

Comparison of the $H$ band detection of
the He{\sc ii} $4686$\AA\ line ({\it e.g.} Soifer {\it et al.}
1995) with the $1640$\AA\ line suggests extinctions
$A_{\rm V}\sim1.3$. 
However, our photoionisation models above
found the Fe{\sc ii}$4658$\AA\ and/or Ar{\sc iv}$4720$\AA\ lines can
contribute up to three times the flux of the He{\sc ii}$4686$\AA\ line
in densities of $10^{10}-10^{12}$ m$^{-3}$, not unreasonable for AGN
narrow line regions. Moreover, 
the Soifer et al. $H-$band spectrum
has very low spectral resolution, and the HeII $4686$\AA\ line flux must be 
treated with some caution. 

This low reddening 
in the rest-frame UV is in marked contrast to the high (\mbox{$A_{\rm
V}\ge5$}) rest frame optical extinction derived from Balmer decrements
({\it e.g.} Soifer {\it et al.} 1995), which led several authors to
infer that at least two 
physically distinct regions contribute to the observed emission line
spectrum ({\it e.g.} Elston {\it et al.} 1994, Soifer {\it et al.}
1995). The limit on the Balmer decrement of
H$\alpha$:H$\beta\ge20$ 
reported by Elston {\it et al.} (1994) is at least in part resolved by
our $K-$band spectrum. The narrower H$\alpha$ component contributes
$\sim25\%$ of the total H$\alpha$ flux,; furthermore our total
H$\alpha$ flux is $\sim25\%$ lower than that reported by Elston {\it
et al.}, though this is within their stated photometric errors. 
However, this still yields
$A_V>4$ in
the frame of F10214+4724, or $A_V>6$ in that of the lens. (Any
$<200$ km s$^{-1}$ H$\beta$ leads to even higher $A_V$.) 
Alternatively, if we assume the Iwamuro {\it
et al.} (1995) H$\beta$ detection is correct,
the data are consistent with zero extinction. 
Further near-infrared
spectroscopy is required to resolve this issue.
We conclude that as yet there is no conclusive evidence for any
significant reddening of the Seyfert-II nucleus.

\section{Starburst properties}

Our limit on the width of the narrowest H$\alpha$ 
component in F10214+4724 is suggestive of a
star-forming region as has previously been suggested by
Kroker et al. (1996). Moreover, 
the fluxes of the [O{\sc i}] $6300$\AA\ and [S{\sc ii}] 
$6724$\AA\ lines (Soifer {\it et al.} 1995) relative to the
narrowest H$\alpha$ 
component are similar to both those seen in spectra of local starburst
galaxies ({\it e.g.} De Robertis \& Shaw 1988) and in high luminosity
IRAS galaxies ({\it e.g.} Leech {\it et al.} 1989). 
We can obtain an estimate of the star formation rate $R$ from the
narrower H$\alpha$ line flux $S$, using the following expression
adapted from Kennicutt (1983):
\begin{equation}
R(z=2.286) = \frac{S(H\alpha) h_{50}^{-2}}{3\times10^{-21}~{\rm
W~m^{-2}}} ~M_\odot~{\rm yr^{-1}}
\end{equation}
We assume a Hubble constant of $H_0=50 ~ h_{50}$ 
km \nolinebreak[4]s$^{-1}$ Mpc$^{-1}$ with $\Omega_0=1$,
$\Omega_\Lambda=0$. 
If we assume the starburst region undergoes a net gravitational 
magnification of ${\cal M}\stackrel{<}{_\sim}10$ (Downes {\it et al.}
1995, Green \& Rowan-Robinson 1996, Graham {\it et al.} 1995), 
and that aperture corrections are negligable, 
we obtain  a star formation rate of 
$\stackrel{>}{_\sim}20 h_{50}^{-2} M_\odot$ per year. This corresponds to
a starburst luminosity of
$\stackrel{>}{_\sim}2.6\times10^{11}h_{50}^{-2}L_\odot$ ({\it e.g.}
Moorwood 1996), {\it i.e.}
$\sim 0.5\%$ of the magnification-corrected bolometric power
output. 

Reddening by intervening dust would further increase this esimate of
the star formation rate. 
Radiative transfer models of F10214+4724 (Green \& Rowan-Robinson
1996) suggest a much larger starburst bolometric fraction of
$\sim0.2-0.4$, with a high starburst UV optical depth of $\tau_{\rm
UV}\sim 800$, {\it i.e.} an $A_{\rm V}\sim 160$. 
Assuming this dust is well mixed with the H$\alpha$ emitting gas ({\it
e.g.} Thronson {\it et al.} 1990), the extinction and star
formation rate is quite
consistent with our observed narrow H$\alpha$ flux, which for this
$A_{\rm V}$ yields a
starburst bolometric fraction of $\sim 0.5$. In both NGC1068
(Green \& Rowan-Robinbson 1996) and F10214+4724, the
starburst and active nuclei appear to make comparable contributions to
the bolometric power output.

\section{Concluding remarks}
The presence of the optically thick O{\sc vi} $1032,1037$\AA\ doublet
appears at face value to imply extremely high densities for narrow
emission line gas,  
$n_{\rm H}\stackrel{>}{_\sim}10^{17}$ m$^{-3}$. However, we argue that
it is more easily attributable to the Ly$\beta$ damping wings of
resonant scattering material, probably withing the AGN narrow line
region. 
%Besides this O{\sc vi}, we make
%two further new identifications: Mg{\sc vi} $1806$\AA\ (ionisation
%potential 141 keV) and (tentatively) [Na{\sc v}]
%$2067.9$,$2069.8$\AA. 
Differential magnification
appears to play a significant role in the emission line spectrum of
F10214+4724, in which
we identify three distinct kinematic components: quasar broad lines
($\sim 4000$ km s$^{-1}$), Seyfert II narrow lines ($\sim 1000$ km
s$^{-1}$) and a starburst component ($\stackrel{<}{_\sim}200$ km
s$^{-1}$). The density, ionisation parameter, number and total mass of
Seyfert II 
narrow line clouds all resemble local Seyferts, and our interpretation
of the 
resonant scattering agrees with that of Villar-Martin {\it et al.}
(1996) for high-redshift radiogalaxies. 
The flux from the narrowest 
H$\alpha$ component is in excellent agreement with radiative transfer
models which comprise similar 
quasar and starburst bolometric contributions.

\section*{Acknowledgements}
\label{sec:acknowledgements}

We thank Carlos Martin for assisting the observations.
The WHT is operated on the island of La Palma by
the Royal Greenwich Observatory in the Spanish Observatorio del
Roque de los Muchachos of the Instituto de Astrofisica de
Canarias. We thank Steve Eales for performing the GASP astrometry
of the $\rm F10214+4724$ field, and for (unwittingly) contributing
observing time to the project.
We also thank Tony Lynas-Gray, Geoff
Smith and Steve Warren
for useful discussions.

\section{Appendix A: Companion galaxies}
\label{sec:appendixa}
Figure A1 shows the spectrum of a galaxy $\sim19''$
south-west of the IRAS galaxy, which fortuitously lay on the slit in
our WHT observations. The feature at around $5510$\AA\ is a probably a
low energy cosmic ray event, but the emission lines at $7155$\AA\ and
(marginally) at $5323$\AA\ appear to be real, since they occur in the
spectra at both nights. 
Cross-correlating this spectrum with old galaxy models
(a 1-Gyr burst aged by $6\times10^8$ years to $1$~Gyr) from Bruzual \&
Charlot (1993) yields two
redshift estimates, $0.428$ and $0.476$ (both $\pm\sim5\%$)
We prefer the former, which identifies the weak emission features as
[O{\sc ii}]$3727$ and [O{\sc iii}]$5007$. 
Close {\it et al.} (1995) report a 
similar redshift, \mbox{$0.429\pm0.002$} for a further galaxy
$\sim23''$ south-west of the IRAS galaxy. 
The UV upturn in figure A1 may indicate a
recent starburst, as expected perhaps if the galaxies are tidally
interacting.
However, there could be a slight systematic error in the redshift of the
Close {\it et al.} companion.
The emission line wavelengths
in their spectrum of F10214+4724 are offset by
a factor $\sim0.99$
compared to other published spectra 
({\it e.g.} Goodrich {\it et al.} 1995, Rowan-Robinson {\it et al.}
1993), and the residuals from the $5577$\AA\ sky feature also 
appear to be offset by a similar amount. 
\vspace*{0.5cm}

\hspace*{0cm}
\vspace*{5cm}
\begin{picture}(100,100)(10,10)
\centering
\put(-30,125){\includegraphics{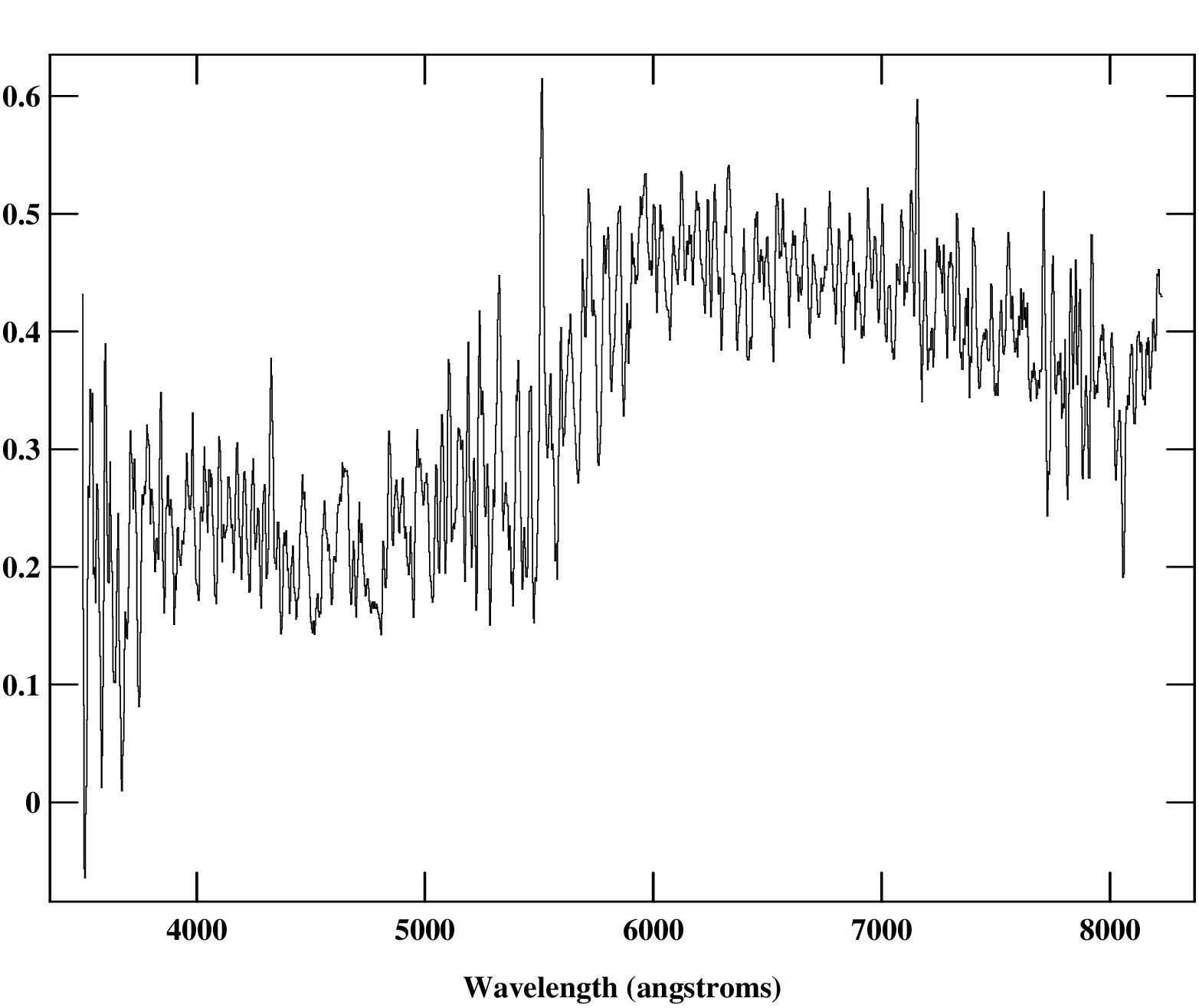}}
\parbox{70mm}{
{\small\vspace*{6cm}
{\bf Figure A1.} 
Serendipitous galaxy $\sim19''$ south-west of F10214+4724. The flux
has units $10^{-20}$ W~m$^{-2}$~\AA$^{-1}$, and the spectrum is
extracted with a full width zero intensity aperture. 
}}
\end{picture}
%\vspace*{-1.5cm}

\begin{figure*}
%\vspace*{5.0cm}
\centering
  \ForceWidth{4.0in}
%  \vSlide{-2cm}
  \TrimTop{5cm}
  \hSlide{-1.5cm}
  \BoxedEPSF{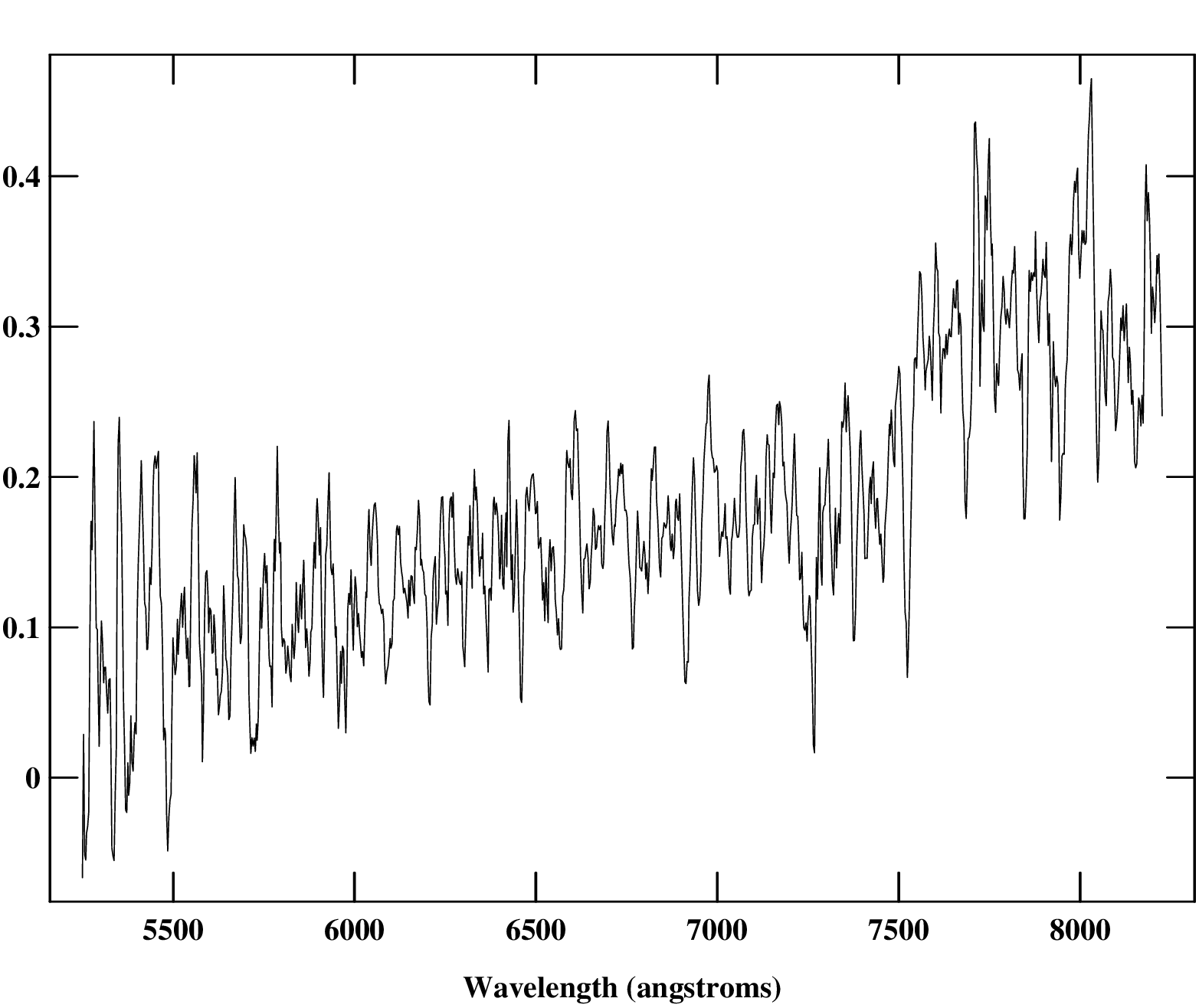}
\parbox{150mm}{
{\bf Figure A2.} 
The sum of the spectra of sources 2 and 3 (in the notation of Matthews
{\it et al.} 1994). These spectra were extracted with 3~pixel
($\sim1.1''$) 
apertures, differing slightly to the method of Serjeant {\it et al.}
1995, though with very little quantitative difference. Source~2
is corrected for contamination by the IRAS galaxy by subtracting a
spectrum of F10214+4724 scaled by the strength of the C{\sc iii}]$1909$
line. The sum has been smoothed with a $5$-pixel ($\sim14$\AA)
boxcar. The flux
has units $10^{-20}$ W~m$^{-2}$~\AA$^{-1}$. 
}
\end{figure*}

The tentative lens redshift of $z\simeq 0.90$ (Serjeant {\it et al.}
1995) is consistent with both
its velocity dispersion ({\it e.g.} Broadhurst \& L\'{e}har 1995) and
surface brightness profile (Eisenhardt {\it et al.} 1996). 
Further support for this lens redshift comes from
figure A2, in which the spectra of sources 2 and 3 (Serjeant {\it et
al.} 1995) are summed. Source 2 is the lensing galaxy and source 3 is
the companion $\sim3.3''$ north east of the IRAS galaxy, in the
notation of Matthews {\it et al.} (1994). If the galaxies are indeed
associated, as 
suggested by their similar spectral energy distributions and tentative
$4000$\AA\ breaks (Serjeant {\it et al.} 1995), as well as hints of
tidal interaction in HST imaging (Eisenhardt {\it et al.} 1995), then
the summed spectrum should show a $\sim\sqrt{2}$ improvement in
signal-to-noise, which is indeed the case in figure A2. 

Several groups have also noted the slight overdensity of galaxies 
in the vicinity of F10214+4724. 
The discovery of a Mg{\sc ii} absorber in the Keck spectrum at
\mbox{$z=1.316$} (Goodrich {\it et al.} 1995), and the possibly
associated pair at $z\simeq 0.4$, 
implies that several
physically independent systems, at a variety of redshifts, contribute
to the overdensity in the field. 

\section{Appendix B: The mystery line at 2067\AA}
\label{sec:appendixb}

In this section we summarise our attempts to identify the 
line at 2067\AA in the rest-frame spectrum of F10214+4724.
We have made an extensive literature search, and investigated 
synthetic spectra generated by the
{\sc cloudy} photoionisation code over wide ranges of  
ionisation parameter and abundances
(v84.12, Ferland 1995; and see section\ref{sec:cloudy}).
This led to only two possible identifications for this line. 
The first of these is the resonant boron 
BIII 2068 doublet, the next resonance line
in the Lithium isoelectronic sequence below the bright OVI, NV and CIV
lines. The tiny cosmic abundance of boron means that this line is
predicted to be immeasurably faint, and we therefore reject this
possibility. The second possibility is the forbidden [NaV] 2067.9/2069.8 
doublet (Mendoza 1983) - a line which is not found in the
{\sc cloudy} v84.12 output.

To determine whether the [NaV] 2067.9/2069.8 doublet
is a plausible identification for the mystery line 
we have compared its predicted strength with 
that of [Ne IV] 2422.5/2425.1. These two pairs of collisionally-excited 
lines arise from electronic transitions within
adjacent ions in the Nitrogen isoelectronic
sequence. In the low-density limit the
ratio of the cooling rates in these doublets is given by

\begin{equation}
\frac{L_{2069}}{L_{2424}}
= \frac{n_{Na{\sc v}}}{n_{Ne{\sc iv}}}\times
  \frac{e^{\frac{-\chi_{2069}}{k T}}}{e^{\frac{-\chi_{2424}}{k T}}}
  \times
  \frac{\Omega_{2069}}{\Omega_{2424}}
  \times
  \frac{2424}{2069}
\end{equation}

(Osterbrock 1989), where $\chi_{\lambda}$ is the exitation potential 
of the upper energy level(s) and $\Omega_{\lambda}$ is the effective
collision strength (see Mendoza 1983). 

Adopting $T = 10^{4} ~ K$ and solar
abundances, and assuming that the fraction of Na in the
Na$^{4+}$ state is similar to the fraction of Ne in the
Ne$^{3+}$ state, the value of this ratio is $\approx  0.005$. The measured
ratio is $\approx 0.23$ so at first sight the putative 
[Na{\sc v}] 2067.9/2069.8 doublet would need to be anomalously strong.
However, at higher densities, collisional de-excitation may become
important. The critical densities for collisional de-excitation $n_{crit}$
are $\sim 10^{11} ~ \rm m^{-3}$ and $\sim 10^{12} ~ \rm m^{-3}$ for the
lower wavelength lines in the Ne and Na doublets respectively
(in each case the higher wavelength doublet line has an order of magnitude
lower $n_{crit}$ than its partner). Given the weakness of the
[OII]3727 line, with $n_{crit} \sim 10^{10} ~ \rm m^{-3}$, in 
F10214+4724, and the strengths of the
[OIII]4959/5007 line and forbidden neon lines 
(with $n_{crit} \stackrel{>}{_\sim} 10^{12} ~ \rm m^{-3}$) 
(Soifer et al. 1995) it is 
certainly feasible that collisional de-excitation can account for the
strength of the putative [Na{\sc v}] doublet with respect to the
[Ne{\sc iv}] doublet.

We conclude that the [Na{\sc v}]2067.9/2069.8 doublet is a 
probable identification for the mystery line in the 
spectrum of F10214+4724, and, by analogy, in NGC1068 and
other active galaxies.


\begin{thebibliography}{99}

\bibitem{100} Adams, T., 1972, MNRAS 174, 439

\bibitem{200} Barvainis R., Tacconi L., Antonucci R., Allion D., Coleman P., 
1994, Nat, 371, 586

\bibitem{250} Bechtold, J., 1994, ApJ Suppl., 91, 1

\bibitem{300}Blandford, R.D., Kochanek, C.S., 1987, ApJ, 321, 658

\bibitem{350}Bland-Hawthorn, J., Sokolowski, J., Cecil, G.,
ApJ 375, 78

\bibitem{400} Broadhurst T. \& Leh\'{a}r J., 1995, ApJ Lett, 450, 41

\bibitem{500}Carico, D.P., Sanders, D.B., Soifer, B.T., Matthews, K.,
Neugebauer, G., 1988, ApJ, 100, 70

\bibitem{525} Charlot, S., Fall, S.M., 1991, ApJ 378, 471

\bibitem{550} Clements, D., van der Werf, P., Krabbe, A., Blietz, M.,
Genzel, R., Ward, M., 1993, Mon. Not. R. astr. Soc., 262, 23P

\bibitem{560} Condon, J.J., Anderson, M.L., Helou, G., 1991, ApJ, 376,
95 

\bibitem{570} Condon, J.J., Huang, Z.-P., Yin, Q.F., Thuan, T.X., 1991,
ApJ, 378, 65 

\bibitem{580} Davidson, K., Netzer, H., 1979, Rev. Mod. Phys., {\bf
51}, 715

\bibitem{590} De Robertis, M.M., Shaw, R.A., 1988, ApJ, 329, 629

\bibitem{600} Diaz, A.I., Prieto, M.A., Wamsteker, W., 1988, A \& A,
195, 53

\bibitem{610} Downes, D., Radford, S.J.E., Greve, A., Thum, C., Solomon, P.M.,
Wink, J.E., 1992, ApJ, 398, L25

\bibitem{650} Downes, D., Solomon, P.M., Radford, S.J.E., 1995, ApJ,
453, 65

\bibitem{700} Dunlop, J.S., Hughes, D.H., Rawlings, S., Eales, S.A., 
Ward, M.J., 1994, Nat, 370, 347

\bibitem{750} Eisenhardt, P.R., Armus, L., Hogg, D.W., Soifer, B.T.,
Neugebauer, G., Werner, M.W., 1996, ApJ, 461, 72

\bibitem{800} Elston, R., McCarthy, P.J., Eisenhardt, P., Dickinson, M.,
Spinrad, H., Januzzi, B.T., Maloney, P., 1994, AJ, 107, 910

%%%\bibitem{900} Faber S.M., Jackson R.E., 1976, ApJ, 204, 668

\bibitem{910} Ferland, G.J., 1993, University of Kentucky Department of
Physics and Astronomy Internal Report
 
\bibitem{920} Ferland, G.J., 1996, {\it Hazy, a Brief Introduction to
Cloudy}, University of Kentucky Department of
Physics and Astronomy Internal Report
 

\bibitem{930} Francis, P., Hewett, P., Foltz, C., Chaffee, F., Weymann,
R., Morris, S., 1991, ApJ, 373, 465

\bibitem{950} Giallongo, E., Cristiani, S., D'Odorico, S., 
Fontana, A., Savaglio, S., 1996, ApJ 466,46

\bibitem{970} Goodrich, R., Miller, J., Martel, A., Cohen, M., Tran, H.,
1996, ApJ., 456, L9

\bibitem{990} Graham, J.R. \& Liu. M.C., 1995, ApJ, in press

\bibitem{1000} Green, S.M., Rowan-Robinson, M., 1996, MNRAS 279, 884


\bibitem{1005} Haehnelt, M.G., Steinmetz, M., Rauch, M., 1997, preprint,
submitted to ApJ

\bibitem{1010} Hines, D.C., Schmidt, G.D., Cutri, R.M., Low, F.J.,
1995, ApJ Lett, 450, 1

\bibitem{1020} Iwamuro, F., Maihara, T., Tsukamoto, H., Oya, S., Hall,
D.B., Cowie, L.L., 1995, PASJ, 47, 265

\bibitem{1030} Jannuzi , B.T., Elston, R., Schmidt, G.D., Smith, P.S.,
Stockman, H.S., 1994, ApJ, 429, L49

\bibitem{1040} Kennicutt, R.C., Jr., 1983, ApJ, 272, 54

\bibitem{1050} Kennicutt, R.C., Jr., Tamblyn, P., Congdon, C.E., 1995,
ApJ 435, 22

\bibitem{1070} Kriss, G.A., {\it et al.}, 1992a, ApJ, 392, 485

\bibitem{1080} Kriss, G.A., {\it et al.}, 1992b, ApJ Lett, 394, 37

\bibitem{1090} Kroker, H., Genzel, R., Krabbe, A., Tacconi-Garman,
L.E., Tecza, M., Thatte, N., 1996, ApJL, 463, 55

\bibitem{1100} Langston G.I., Conner S.R., Leh\'{a}r J., Burke B.F.,
Weiler K.W.,  
Nat, 1990, 344, 43

\bibitem{1200} Larkin J.E., et al., 1994, ApJ, 420, L9

\bibitem{1250} Lawrence, A., Rowan-Robinson, M., Oliver, S., Taylor,
A., McMahon, R.G., Broadhurst, T., Scarrot, S.M., Rolph, C.D., Draper,
P.W., Ellis, R.S., Tadhunter, C., Condon, J.J., Lonsdale, C.J.,
Hacking, P., Conrow, T., Efstathiou, G.P., Saunders, W.S., 1993,
MNRAS, 260, 268

\bibitem{1270} Leech, K.J., Penston, M.V., Terlevich, R., Lawrence, A.,
Rowan-Robinson, M., Crawford, C., 1989, MNRAS 240, 349

\bibitem{1300} Matthews K., et al., 1994,
ApJ, 420, L13 

\bibitem{1350} Moorwood, A.F.M., 1996, Space Science Reviews. 77, 303

\bibitem{1400} Mountain, C.M., Robertson, D.J., Lee, T.J., Wade, R.,
1990, Instumentation in Astronomy {\sc vii}, ed. D.L. Crawford
(Proc. SPIE, 1235), 25

\bibitem{1425} Narayan R., Wallington S., 1992, in Kayser R., Schramm T., Nieser
L., eds, Gravitational Lenses. Springer-Verlag, Berlin, p.\ 12

\bibitem{1450} Nadeau , D., Yee, H.K.C., Forrest, W.J., Garnett, J.D.,
Ninkov, Z., Pipher, J., 1991, ApJ, 376, 430

\bibitem{1470} Nelson, C.H., Whittle, M., 1996, ApJ., 465, 96

\bibitem{1500} Neufeld, D.A., McKee, C.F., 1988, ApJ, 331, L87
 
\bibitem{1510} Osterbrock, D.E., 1989, {\it ``Astrophysics of Gaseous
Nebulae and Active Galactic Nuclei''}, University Science
Publications, Mill Valley, CA, USA

\bibitem{1520} Osterbrock, D.E., 1993, ApJ 404, 551

\bibitem{1550} Pei, Y., 1992, ApJ 395, 130

\bibitem{1570} Phinney, E.S., 1989, Theory of Accretion Disks, W. Dushl,
F. Meyer and J. Frank, eds, (Dordrecht: Kluwer Academic Publishers) pp
457-470 

\bibitem{1580} Radford, S.J.E., Brown, R.L., Vanden Bout, P.A., 1993, A
\& A, 271, L21

\bibitem{1590} Ramsay, S.K., Mountain, C.M., Geballle, T.R., 1992,
MNRAS, 259, 751

\bibitem{1600} Rocca-Volmerange B., Guiderdoni B., 1988, A\& AS 75, 93

\bibitem{1620} Rowan-Robinson, M., 1995, MNRAS, 272, 737

\bibitem{1650} Rowan-Robinson, M., Crawford, C., 1989, MNRAS, 238, 523

\bibitem{1700} Rowan-Robinson, M., et al., 1991, Nat, 351, 719

\bibitem{1800} Rowan-Robinson, M., et al., 1993, MNRAS, 261, 513

%%%\bibitem{1900} Schneider P., Ehlers J. \& Falcko, E.E., 1992,
%%%Gravitational Lenses, Springer-Verlag, Berlin.

\bibitem{1950} Scoville, N.Z., Yun, M.S., Brown, R.L., Vanden Bout,
P.A., 1995, ApJ., 449, L109

\bibitem{1960} Serjeant, S., Lacy, M., Rawlings, S., King,, L.J.,
Clements, D.L., 1995, MNRAS 276, L31

\bibitem{1970} Snidjers, M.A.J., Netzer, H., Boksenberg, A., 1986,
MNRAS, 222, 549

\bibitem{2000} Soifer, B.T., Cohen, L., Armus, K., Matthews G., 
Neugebauer G., Oke J.B., 1995, ApJ, 443, L65

\bibitem{2100} Soifer, B.T., Neugebauer, G., Matthews, K., 
Lawrence, C., and Mazzarella, J., 1992, ApJ, 399, L55

\bibitem{2200} Solomon, P.M., Downes, D., Radford, S.J.E., 1992, ApJ, 398, L29

\bibitem{2220} Sopp, H.M., Alexander, P., 1991, MNRAS, 251, 14P

\bibitem{2230} Steidel, C.C., Giavalisco, M., Pettini, M., Dickinson, M.,
Adelberger, K.L., 1996, ApJ Lett, 462, 17

\bibitem{2240} Thronson, H.A., Majewski, S., Descartes, L., Hereld, M.,
1990, ApJ 364, 456

\bibitem{2250} Tran, H., 1995, ApJ, 440, 597

\bibitem{2270} Trentham, N., 1995, MNRAS, 277, 616

\bibitem{2300} Turner E.L., Ostriker J.P., Gott J.R., 1984, ApJ,, 284, 1

\bibitem{2310} Villar-Martin, M., Binette, L., Fosbury, R.A.E., 1996, 
A. \& A. 312, 751

\bibitem{2320} Wolfe, A.M., Lanzetta, K.M., Foltz, C.B., Chaffee, F.H.,
1995, ApJ 454, 698

\bibitem{2400} Warren, S., Hewett, P., Osmer, P., 1994, ApJ, 421, 412

\bibitem{2500} Wilson, C.D., Walker, C.E., Thornley, M.D., 1997, ApJ 482,
131 

\end{thebibliography}
\end{document}